\documentclass[aps,prd,twocolumn,superscriptaddress,preprintnumbers,floatfix,nofootinbib,amsmath,amssymb]{revtex4-2}

\usepackage{graphicx}
\usepackage{dcolumn}
\usepackage{bm}
\usepackage{hyperref}
\usepackage{orcidlink}
\usepackage{enumitem}

\newcommand{\na}{New Astronomy} 
\newcommand{\mnras}{MNRAS}		
\newcommand{\jcap}{JCAP}		

\newcommand{\physrep}{Phys.~Rep.}   

\begin{document}

\title{Faster than SAM: An empirical model for the tidal evolution of dark matter substructure around strong gravitational lenses}

\author{Xiaolong Du\,\orcidlink{0000-0003-0728-2533}}
\email{xdu@astro.ucla.edu}
\affiliation{Department of Physics and Astronomy, University of California, Los Angeles, California 90095, USA}

\author{Daniel Gilman\,\orcidlink{0000-0002-5116-7287}}
\email{gilmanda@uchicago.edu}
\affiliation{Department of Astronomy $\&$ Astrophysics, University of Chicago, Chicago, Illinois 60637, USA}
\affiliation{Brinson Prize Fellow}

\author{Tommaso Treu\,\orcidlink{0000-0002-8460-0390}}
\affiliation{Department of Physics and Astronomy, University of California, Los Angeles, California 90095, USA}

\author{Andrew Benson\,\orcidlink{0000-0001-5501-6008}}
\affiliation{Carnegie Observatories, 813 Santa Barbara Street, Pasadena, California 91101, USA}

\author{Charles Gannon\,\orcidlink{0009-0009-0443-3181}}
\affiliation{University of California, Merced, 5200 N Lake Road, Merced, California 95341, USA}

\date{\today}

\begin{abstract}
Strong gravitational lenses enable direct inference of halo abundance and internal structure, which in turn enable constraints on the nature of dark matter and the primordial matter power spectrum. However, the density profiles of dark subhalos around the main deflector of a strong lens system also depend on tidal evolution inside the host, complicating the interpretation of strong-lensing inferences. We present a model for subhalo tidal evolution that accurately predicts the bound mass function and the density profiles of tidally stripped subhalos that appear near the Einstein radius of a typical deflector for a variety of dark matter models. This model matches predictions from the semianalytic model (SAM) {\sc galacticus}, but enables the simulation of subhalo populations in seconds, rather than hours. We use this model to examine the expected number of subhalos near the Einstein radius of a typical lens, and examine their lensing signals. We show that in cold dark matter the amplitude of the bound mass function is suppressed by a factor of $20$ relative to the infall mass function, and $87 \%$ of subhalos appearing in projection near the Einstein radius of a typical strong lensing deflector have lost more than $80\%$ of their mass since infall. Tidal stripping becomes increasingly severe in dark matter models with suppressed small-scale power, such as warm dark matter. This model will be used to forward model subhalo populations in forthcoming analyses of strong lens systems.
\end{abstract}

\maketitle

\section{Introduction}
\label{sec:intro}
The next frontier for cosmic probes of dark matter substructure lies on subgalactic scales, in dark-matter halos with masses below $10^8 M_{\odot}$. The abundance and internal structure of these objects, most of which are not expected to contain luminous galaxies \cite{Nadler20}, depend on the particle nature of dark matter \cite{Buckley18}. Several observational techniques, such as the study of stellar streams and strong gravitational lensing have the capability to study dark-matter structure in this regime (for a review, see \cite{DrlicaWagner22}). 

One of the principal challenges associated with the study of dark-matter substructure is the interpretation of inferences related to dark-matter subhalos, objects which have become gravitationally bound to a larger host halo and orbit within its gravitational potential. While many theories make relatively clear predictions for the properties of isolated dark-matter halos in the field, subhalos experience a complex nonlinear evolution in the tidal field of their host. After a subhalo is accreted, it begins losing mass due to tidal stripping, sinks over time to the center of the host through dynamic friction, and heats up through tidal forces exerted by the host. Interpreting inferences of subhalo properties therefore requires a clear understanding of how these processes affect the bound mass and internal structure of subhalos.

Strong gravitational lensing enables a direct, purely gravitational investigation of dark-matter subhalos around, and halos along the line of sight to, distant galaxies (for a review, see \cite{Vegetti:2023mgp}). Strong lensing inferences must also contend with uncertainties related to the tidal evolution of subhalos, and how this process will affect the observable features of these systems, which can include extended luminous arcs or multiple images of a background quasar. However, most analyses of tidal evolution have focused on substructure within the Milky Way, and in particular, how subhalos respond to tidal forces exerted by the Galactic disk \citep[][]{Garrison-Kimmel17,Errani17,WebbBovy20,Wang24}. The situation changes in massive elliptical galaxies, the systems most commonly acting as strong lensing deflectors, because massive elliptical galaxies typically reside in group-scale hosts with virial masses $\sim 10^{13} M_{\odot}$ \cite{Lagattuta10,Auger10}. The tidal evolution of dark substructure within these systems is expected to differ from subhalos around the Milky Way due to their larger physical size and the absence of a stellar disk. 

A powerful way to study the tidal evolution of dark subhalos around group-scale hosts has been through cosmological zoom-in simulations \cite{Fiacconi16,Nadler23,Gannon:2025nhr}. Cosmological zoom-in simulations have also been performed to study the subhalo statistics for halos with different masses, from the Large Magellanic Cloud-size halos to cluster-size ones (see e.g. \cite{Nadler23}). These simulations are usually limited by numerical resolution, which imposes a minimum resolved mass threshold and can introduce unphysical outcomes, for example, artificial disruption~\cite{vandenBosch:2018tyt} (although Ref.~\cite{He:2024hsx} discusses why artificial disruption might not be a severe problem). Moreover, resolving low-mass $\left(m<10^8 M_{\odot}\right)$ halos in N-body simulations, particularly for group-scale systems, is computationally expensive due to the high mass of the host. 

Semianalytic models (SAMs) offer a complementary avenue for studying the nonlinear evolution of cosmic structure~\cite{Taylor:2000zs,Taffoni:2003sr,Pullen:2014gna,Yang:2020aqk}. Rather than dealing with extremely large quantities of particles evolving gravitationally, SAMs track the evolution of derived quantities, such as halo mass, tidal radii, orbital position and velocity, halo density profiles, etc. For this work we use the semianalytic model {\sc galacticus} ~\cite{2012NewA...17..175B}.\footnote{\url{https://github.com/galacticusorg/galacticus}} {\sc galacticus} contains several physical models for processes such as tidal stripping, tidal heating and dynamical friction~\cite{Pullen:2014gna,Yang:2020aqk,Du:2024sbt}, which have been calibrated to high-resolution idealized N-body simulations~\cite{Du:2024sbt}. After calibrating a SAM, one can simulate subhalo populations with much less computing resources, making it a useful way to explore different dark matter models, determine statistical uncertainties, and examine trends across redshift and halo mass that require a large number of simulations, e.g.,~\cite{Gannon:2025nhr}. Furthermore, it has been shown that the artificial effects in cosmological simulations can be modeled and removed from the prediction, thus providing more robust results for subhalo properties, especially in the central region of the host~\cite{Benson:2022tzm}.

However, for some applications, even running SAMs is too slow. For example, when studying dark matter substructure populations with quadruply imaged, one must generate hundreds of thousands or millions of populations of subhalos and line-of-sight halos per lens, e.g.,~\cite{Gilman20,Gilman24}. To remedy this issue, we develop a new tidal stripping model based on {\sc galacticus}, and implement it in the open-source software {\tt{pyHalo}},\footnote{\url{https://github.com/dangilman/pyHalo}} a code which has already been used extensively for substructure lensing analyses in a variety of dark matter models \cite{Laroche22,Gilman23,Keeley24}, as well as millilensing of multiply imaged supernova \cite{Kelly23,Larison25}. In a similar way to how a SAM, such as {\sc galacticus}, incorporates a layer of abstraction to speed up calculations relative to N-body simulations, codes like {\tt{pyHalo}} circumvent some of the more expensive calculations performed with SAMs by generating substructure populations``in place", meaning they do not explicitly model subhalo orbits and other processes occurring over time. Realizations of dark matter substructure are created as they would appear in a strong lensing deflector at the time of lensing, typically in seconds.  

\begin{figure*}
\includegraphics[width=0.9\textwidth]{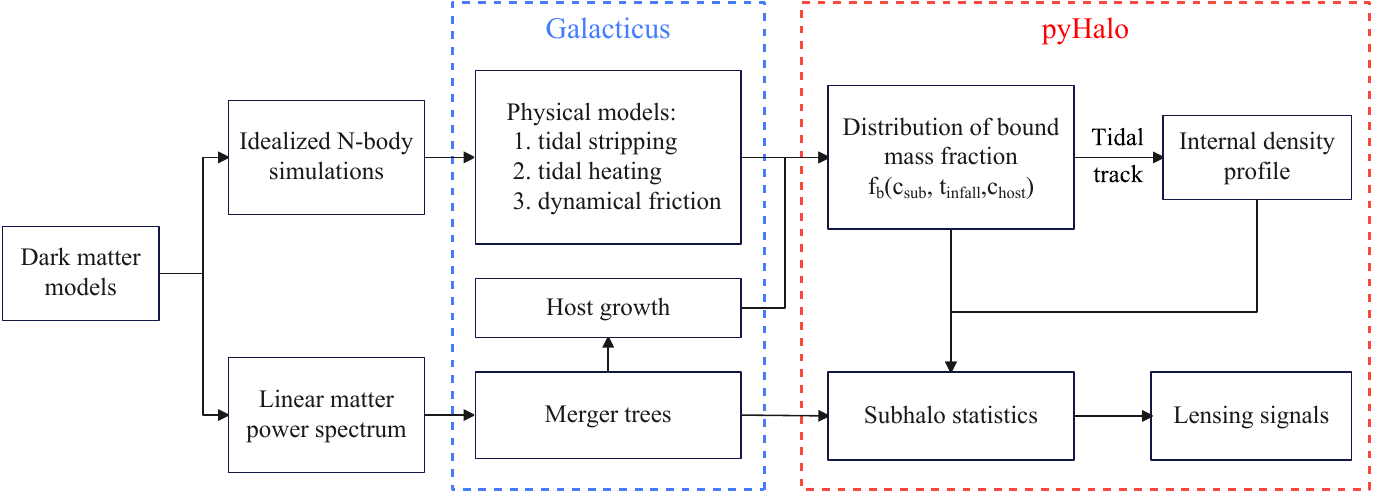}
\caption{Flow chart of a new framework for modeling tidal evolution of subhalos and predicting their lensing signals.}
\label{fig:flow_chart}
\end{figure*}

Figure~\ref{fig:flow_chart} shows a flow chart of our new framework. Given a specific dark matter model, we need some basic information such as the linear matter power spectrum which can be computed using Boltzmann codes like {\sc CAMB}~\cite{Lewis:1999bs} or {\sc CLASS}~\cite{Blas:2011rf}. Knowing the linear matter spectrum, we use the semianalytic code {\sc galacticus} to generate synthetic merger trees based on the extended Press-Schechter (EPS) formulism~\cite{Press:1973iz,Bower:1991kf,Bond:1990iw,Lacey:1993fec,Cole:2000ex,Parkinson:2007yh}. The merger trees contain information about when dark matter halos form and how they merge into more massive structures. {\sc galacticus} then tracks the evolution of subhalos that become subhalos of a larger host, accounting for the various complicated effects that occur post-infall. Based on the results output by {\sc galacticus}, we can implement simplified parametric formulations for the subhalo mass function, subhalo density profiles in {\tt{pyHalo}} for use in forward modeling analyses of strong lens systems.

In this work, we develop a tidal stripping model with {\sc galacticus} that predicts bound mass fraction $f_{\rm{bound}}$ of a subhalo in a group-scale host, conditioned on the subhalo appearing in projection near the Einstein radius of a typical strong lens system, the subhalo's concentration at infall, the time since infall, and the concentration of the host halo. From the bound mass fraction, we then utilize the tidal track, a universal relation between the bound mass fraction of subhalo and its density transfer function $T=\rho(r,t)/\rho(r,0)$~\cite{Hayashi:2002qv,Penarrubia:2007zx,Penarrubia:2010jk,Green:2019zkz,Errani:2020wgn,Du:2024sbt}, to predict the subhalo's internal density profile. We perform the calculations of subhalo evolution for this model using {\sc galacticus}, and then develop an empirical model in {\tt{pyHalo}} that allows us to rapidly simulate populations of dark-matter subhalos. As we will show, the model implemented in {\tt{pyHalo}} yields an excellent representation of subhalo density profiles emerging from the full-physics calculations performed with {\sc galacticus}. Furthermore, the explicit dependence on the subhalo infall concentration allows us to use this model in a variety of cosmological contexts besides cold dark matter (CDM), provided these alternative dark matter scenarios do not affect the evolution of structures on the scale of the host and the tidal evolution is not significantly affected by nongravitational interactions.

This paper is organized as follows: In Sec.~\ref{sec:sub_evo_host}, we review the prescriptions for subhalo tidal evolution implemented in {\sc galacticus} (Sec.~\ref{ssec:tidalstripping}), and then describe how we model the host halo evolution across cosmic time given its properties at $z=0.5$ (Sec.~\ref{ssec:hostevolution}). In Sec.~\ref{sec:infall}, we discuss how we treat the infall time of subhalos, a key ingredient of our model that requires special care due to the fact that some subhalos become subhalos of other halos before becoming subhalos of the main host. In Sec.~\ref{sec:pyhalosims}, we detail an empirical model implemented in {\tt{pyHalo}} that reproduces the properties of tidally evolved subhalo populations in {\sc galacticus}. In Sec.~\ref{ssec:lensingsignatures}, we examine the strong lensing signatures of tidally evolved dark subhalos. We summarize our results and provide concluding remarks in Sec.~\ref{sec:disscussion}. Through this work, for simplicity, we assume a cosmological model with $\Omega_m=0.3153$, $\Omega_{\Lambda}=0.6847$, $\Omega_b=0.04930$, $\sigma_8=0.8111$, and $H_0=67.36\,{\mathrm{km/s/Mpc}}$~\cite{Planck:2018vyg}, although of course our results can be generalized to any set of cosmological parameters.

\section{Subhalo evolution in an evolving host potential}
\label{sec:sub_evo_host}
To study the tidal evolution of subhalos, we follow the semianalytic models presented in~\cite{Du:2024sbt}, which are calibrated to high-resolution idealized $N$-body simulations. In Sec.~\ref{ssec:tidalstripping}, we begin by reviewing key ingredients of the tidal stripping model in {\sc galacticus}. In Sec.~\ref{ssec:hostevolution}, we describe a model for predicting the evolution of a host halo backwards in time, conditioned on it having a certain concentration and a certain mass at a certain redshift (typically $z=0.5$ for a massive lens galaxy). Throughout this section, particularly in Sec.~\ref{ssec:hostevolution}, we will often refer to two kinds of simulations performed with {\sc galacticus}, which we define here for reference: 
\begin{enumerate}[label=(\roman*)]
    \item full-physics model: this is analogous to a cosmological simulation. Merger trees will first be constructed using the EPS formulism. Each node in the tree represents a dark matter halo. Then {\sc galacticus} starts from the leaf nodes and evolves the nodes forward in time and takes into account different kinds of physics such as mass growth, mergers, tidal evolution, etc.
    \item idealized simulations: this is analogous to an idealized N-body simulation. Each simulation contains only one subhalo evolving in a static or evolving host potential.
\end{enumerate}

\subsection{Tidal effects}
\label{ssec:tidalstripping}
As the subhalo evolves in the host potential, it is subjected to three main processes that affect its bound mass and density profile at later times: tidal stripping, tidal heating, and dynamical friction. In this subsection, we review how these processes are implemented in {\sc galacticus}. 

\subsubsection{Tidal stripping}
The mass loss rate of a subhalo is computed as
\begin{equation}
\frac{dm_{\rm sub}}{dt}=-\alpha_\mathrm{s} \frac{m_{\rm sub}-m_{\rm{sub}}(<r_\mathrm{t})}{T_{\rm loss}},
\label{eq:mass_loss}
\end{equation}
where $\alpha_s=3.8$ is the coefficient of tidal stripping strength, $r_\mathrm{t}$ is the tidal radius, and $T_{\rm loss}$ is the timescale for the mass loss. To compute $r_\mathrm{t}$, we solve the equation
\begin{equation}
r_\mathrm{t}=\left(\dfrac{G m_{\mathrm{sub}}(<r_\mathrm{t})}{\gamma_\mathrm{c}\omega^2-\left.\frac{\mathrm{d}^2\Phi}{\mathrm{d}r^2}\right\vert_{r_{\mathrm{sub}}}}\right)^{1/3}.
\label{eq:r_t}
\end{equation}
Here $\gamma_c$ is a coefficient that controls the contribution from centrifugal force and $\Phi$ is the gravitational potential of the host at the subhalo position. Following Ref.~\cite{Du:2024sbt}, we set $\gamma_c=0$. The mass loss timescale is taken to be the dynamical timescale at $r_\mathrm{t}$: $T_{\rm dyn}=2\pi\sqrt{r_\mathrm{t}^3/16 \mathrm{G} m_{\rm sub}(<r_\mathrm{t})}$.

\subsubsection{Tidal heating}
As the subhalo loses its mass, its density profile also evolves due to tidal heating effects. The heating energy obtained by a mass shell with an initial radius of $r_i$ is computed as~\cite{Benson:2022tzm}
\begin{eqnarray}
\Delta \epsilon(r_i) &=& \Delta \epsilon_1 (r_i) + \Delta \epsilon_2 (r_i) \nonumber \\
&=& \Delta \epsilon_1 (r_i) + \sqrt{2} f_2 (1+\chi_v) \sqrt{\Delta \epsilon_1(r_i) \sigma^2_r(r_i)},
\label{eq:heating_improve}
\end{eqnarray}
where $\Delta \epsilon_1$ and $\Delta \epsilon_2$ are the contributions from the first-order and second-order terms, respectively, $f_2=0.554$ is the coefficient of the second-order term, $\chi_{v}=-0.333$ is the position-velocity correlation, and $\sigma_r$ is the radial velocity dispersion of subhalo. The first-order term is computed by integrating the differential equation
\begin{equation}
\Delta\dot{\epsilon}(r) = \frac{\epsilon_\mathrm{h}}{3} \left[1+(\omega_\mathrm{p} T_{\rm shock})^2\right]^{-\gamma_\mathrm{h}} r^2 g_\mathrm{ab} G^\mathrm{ab}.
\label{eq:heating_rate}
\end{equation}
Here, $\Delta\dot{\epsilon}$ denotes the time derivative of $\Delta{\epsilon}$, $\epsilon_\mathrm{h}=0.0732$ is a coefficient, $\omega_\mathrm{p}$ is the angular frequency of particles at the half-mass radius of the subhalo, $T_{\rm shock}=r_{\rm sub}/V_{\rm sub}$ is the timescale of tidal shock, $\gamma_\mathrm{h}$ is the adiabatic index, $g_{ab}$ is the tidal tensor, and $G_{ab}$ is the time integral of $g_{ab}$~\cite{Pullen:2014gna,Yang:2020aqk}:
\begin{equation}
G_{ab}(t)=\int_0^t dt' \left[ g_{ab}(t)-\beta_\mathrm{h}\frac{G_{ab}(t)}{T_{\rm orbit}}\right].
\label{eq:Gab}
\end{equation}
Here we take $\gamma_h=0$ and $\beta_h=0.275$. After obtaining the heating energy, the mass shell will expand to a new radius $r_f$. Assuming virial equilibrium, we have
\begin{equation}
\Delta \epsilon = \frac{\mathrm{G} M_\mathrm{i}}{2 r_\mathrm{i}}-\frac{\mathrm{G} M_\mathrm{i}}{2 r_\mathrm{f}}.
\label{eq:r_i_r_f}
\end{equation}
The density after heating is then computed as
\begin{equation}
\rho_\mathrm{f} = \rho_\mathrm{i} \frac{r_\mathrm{i}^2}{r_\mathrm{f}^2} \frac{\mathrm{d}r_\mathrm{i}}{\mathrm{d}r_\mathrm{f}}.
\label{eq:rho_f}
\end{equation}

\subsubsection{Dynamical friction}
The position of a subhalo is tracked by solving the following equation
\begin{equation}
\frac{d^2\mathbf{x}}{dt^2}=\mathbf{a}_\mathrm{g} + \mathbf{a}_\mathrm{df},
\end{equation}
where $\mathbf{a}_\mathrm{g}$ is the gravitational acceleration from the host and $\mathbf{a}_\mathrm{df}$ is the acceleration due to dynamical friction~\cite{1943ApJ....97..255C}
\begin{eqnarray}
\mathbf{a}_{\rm{df}}=-4\pi \mathrm{G}^2 \ln\Lambda m_{\rm{sub}}\rho_{\rm{host}}(r_{\rm{sub}})\dfrac{\mathbf{V}_{\rm{sub}}}{V_{\rm{sub}}^3}\nonumber \\
\left[{\rm erf}(X_v)-\dfrac{2X_v}{\sqrt{\pi}}\exp\left(-X_v^2\right)\right].
\label{eq:dyn_friction}
\end{eqnarray}
Here $\rho_{\rm host}$ is the host density at the subhalo position, $r_{\rm sub}$ is the distance to the host center, $X_v=V_{\rm{sub}} / \sqrt{2}\sigma_v$ with $V_{\rm{sub}}$ the velocity of subhalo, $\sigma_v$ is the velocity dispersion of host particles, and $\ln\Lambda$ is the Coulomb logarithm. In \cite{Du:2024sbt}, the Coulomb logarithm was not calibrated because the considered subhalo therein is too small to be significantly affected by dynamical friction. In this work, we compare the results from the semianalytic models with the Symphony cosmological zoom-in simulations~\cite{Nadler23} and find that $\ln\Lambda=4$ gives subhalo halo mass function that is in good agreement with cosmological simulations. Note that although we focus on small subhalos with $m_{\mathrm{sub}}/M_{\mathrm{host}}\lesssim 10^{-3}$ at the lens redshift in this work, the mass ratio might be larger at the infall time of subhalos. So the dynamical friction effect is not negligible.

\subsection{Conditional evolution of the host halo}
\label{ssec:hostevolution}
In \cite{Du:2024sbt}, when calibrating the model parameters, a static host (fixed $\rho_0$ and $r_{\rm s}$) was assumed. In reality, the host will grow as it accretes mass from subhalos and its environment, and therefore the tidal field seen by subhalos changes with time. In particular, at early times, when the host halo is smaller, subhalos are accreted closer to the host center leading to stronger tidal forces.

The task of predicting the properties of dark matter halos over cosmic time is typically solved in the forward direction---given the mass accretion history (MAH) of a halo, one predicts its mass and concentration at some redshift of interest. From an observational standpoint, however, we only have access to information about the host halo properties at some later time, which for a typical lens system corresponds to $z \sim 0.5$. We will therefore need to solve this problem in the backwards direction, and model the evolution of the host at a redshift $z$ given $M_{200} = 10^{13} M_{\odot}$ and a concentration $c_{\rm{host}}$ at $z_0 = 0.5$.

To begin, we assume the host has a Navarro-Frenk-White (NFW) profile~\cite{Navarro:1996gj},
\begin{equation}
\rho(r)=\frac{\rho_0}{\frac{r}{r_{\rm s}}\left(1+\frac{r}{r_{\rm s}}\right)^2},
\label{eq:NFW}
\end{equation}
where $\rho_0$ is characteristic density and $r_{\rm s}$ is the scale radius. The host density profile is determined by its virial mass $M_{200}$ and concentration parameter $c_{200}=R_{200}/r_{\rm s}$. Throughout this paper, we define the virial radius $R_{200}$ as the radius enclosing a mean density of $200$ times of the critical density of the Universe.

The MAHs and concentrations of dark matter halos have been studied extensively with both numerical simulations and semianalytic models, which suggests that the halo concentrations are closely correlated with the MAHs~\cite{Navarro:1996gj,Navarro:1995iw,Bullock:1999he,Wechsler:2001cs,Zhao:2003jf,Ludlow:2013vxa,Ludlow:2016ifl,Diemer:2018vmz,Johnson:2020ufy,Lopez-Cano:2022vza}. For example, in Ref.~\cite{Ludlow:2013vxa}, it was found that if we rescale the halo mass and the critical density of the Universe at redshift $z$ by their current values, these two quantities can be described also by an NFW profile with a concentration of $c_{\mathrm{MAH}}$: $M_{200}(z)/M_{200,0}$ is analogous to the halo mass enclosed within a radius of $r$ and $\rho_{\mathrm{crit}}(z)/\rho_{\mathrm{crit},0}$ is analogous to the mean density within $r$. $c_{\mathrm{MAH}}$ is related to the halo concentration at the current time by \citep{Ludlow:2013vxa}
\begin{equation}
c_{200} = 2.9 (1 + 0.614 \, c_{\mathrm{MAH}})^{0.995}.
\label{eq:c_MAH}
\end{equation}
Therefore, given the concentration of a halo, its MAH can also be determined.

\begin{figure*}
\includegraphics[width=0.45\textwidth]{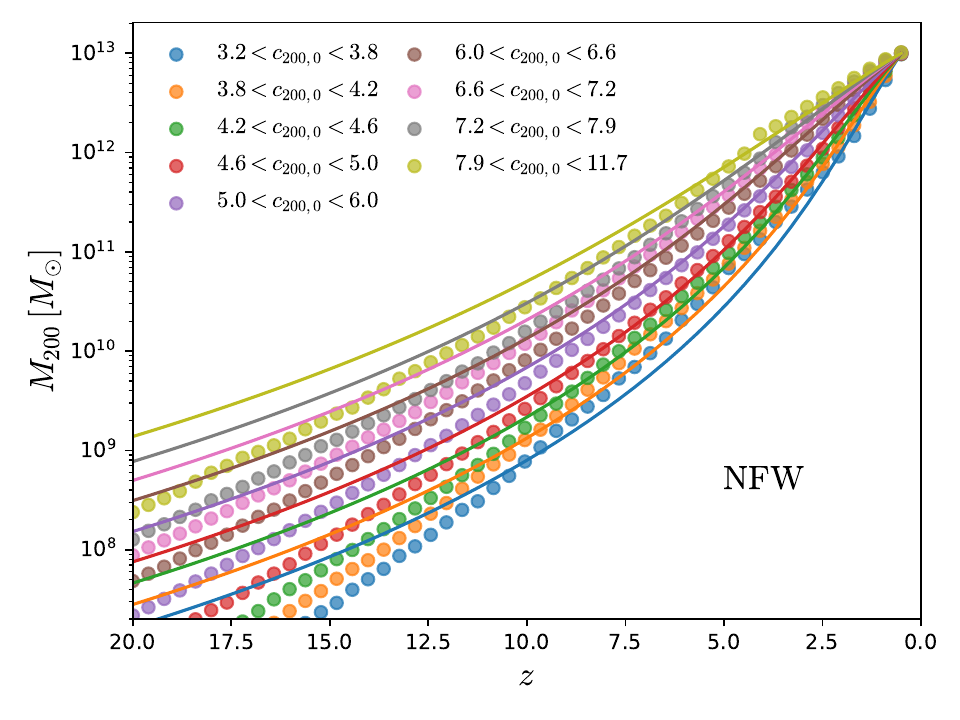}
\includegraphics[width=0.45\textwidth]{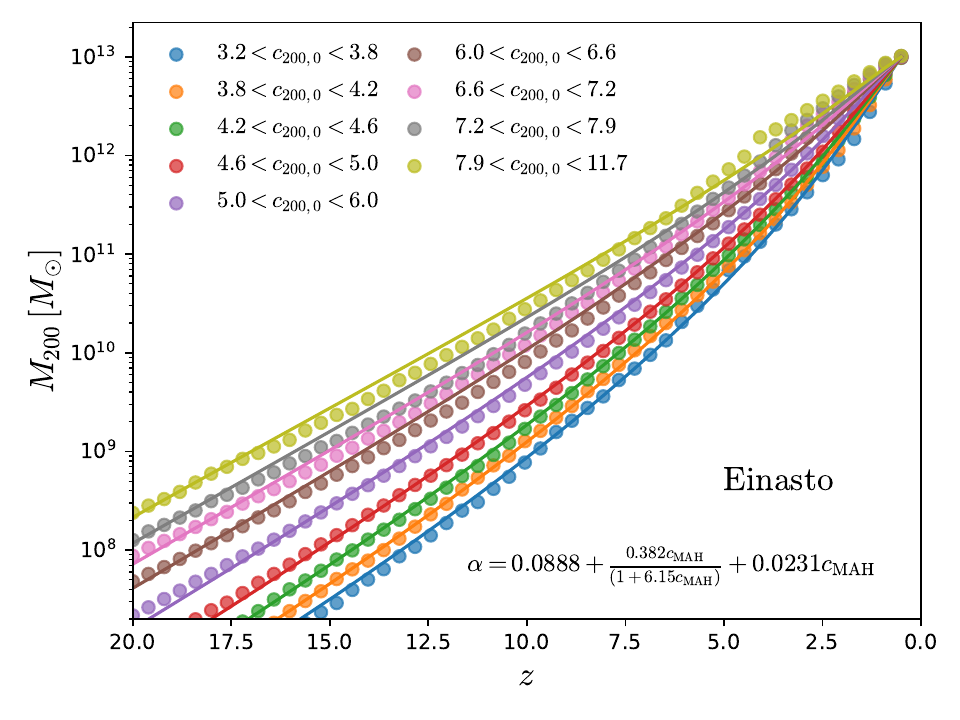}
\caption{Mass accretion history of the hosts binned by concentrations at the current time ($z_0=0.5$). The host mass is fixed at $10^{13} M_{\odot}$ at $z_0=0.5$. The filled circles show the results derived from merger trees. The solid curves show the model predictions: in the left panel, the rescaled mass accretion history is assumed to follow an NFW profile as in Ref.~\cite{Ludlow:2013vxa}; in the right panel, the NFW profile is replaced by an Einasto profile.}
\label{fig:MAHs}
\end{figure*}

To check how well the model works, we generate $16,384$ merger trees using {\sc galacticus} with a mass resolution of $M_{\mathrm{res}}=10^7 M_{\odot}$. The root halo has a mass of $10^{13}M_{\odot}$ at $z_0=0.5$. The merger rates are taken from Ref.~\cite{2019MNRAS.485.5010B}. We then track the mass of the main progenitor along the trees (the main branch). The halo concentration is also determined from the merger trees following the model presented in Ref.~\cite{2019MNRAS.485.5010B} which has been shown to correctly reproduce the relation between the halo concentration and MAH~\cite{Ludlow:2016ifl}. The MAHs of the halos binned by their concentration at the current time are shown in Fig.~\ref{fig:MAHs} (filled circles). As a comparison, we also show the prediction from the Ludlow model (solid curves in the left panel). As can be seen, the model predictions agree well with results from merger trees at $z\lesssim 10$. At higher redshifts, the model tends to overestimate the halo mass. As shown in many previous studies, the halo density profiles in cosmological simulations are better described by the Einasto profile~\cite{1965TrAlm...5...87E,Navarro:2003ew},
\begin{equation}
\rho_{\mathrm{Einasto}} = \rho_{-2} \exp\left[-\frac{2}{\alpha}\left(\left(\frac{r}{r_{-2}}\right)^{\alpha}-1\right)\right],
\label{eq:Einasto}
\end{equation}
where $r_{-2}$ is the radius at which the logarithmic slope is $-2$ (equivalent to $r_{\rm s}$ in the NFW profile), $\rho_{-2}$ is the density at $r_{-2}$, and $\alpha$ controls the radial dependence of the logarithmic slope. This motivates us to use the Einasto profile for determining the MAH as in Ref.~\cite{Ludlow:2013bd}. We find that if we allow $\alpha$ to vary with $c_{\mathrm{MAH}}^{\rm E}$ as
\begin{equation}
\alpha_{\mathrm{MAH}}^{\rm E} = 0.0888 + \frac{0.382 c_{\mathrm{MAH}}}{(1 + 6.15 c_{\mathrm{MAH}})}+ 0.0231 c_{\mathrm{MAH}},
\label{eq:alpha}
\end{equation}
we obtain a much better fit to the results from merger trees---see the solid curve in the right panel of Fig.~\ref{fig:MAHs}. For the cases we consider, the value of $\alpha_{\mathrm{MAH}}$ is $0.12 \sim 0.2$, which is broadly consistent with that found in simulations for the halo profiles~\cite{Gao:2007gh}. But we emphasize that $\alpha_{\mathrm{MAH}}$ is not necessarily equal to the $\alpha$ parameter in the halo profile. Nevertheless, with the Einasto profile, the predicted host mass is similar to that derived using an NFW profile at low redshifts, but the prediction is lower at very high redshifts. This is expected given that the Einasto profile is shallower than the NFW profile in the inner region. As we switch to the Einasto profile, we find that best-fit $c_{\mathrm{MAH}}^{\rm E}$ is in general lower than that assuming an NFW profile, which can be well fitted by
\begin{equation}
c_{200} = 3.08 + \frac{5.46 c_{\mathrm{MAH}}}{(1 + 2.89 c_{\mathrm{MAH}})} + 1.61 c_{\mathrm{MAH}}.
\label{eq:c_MAH_Einasto}
\end{equation}
We have verified that Eqs.~\eqref{eq:alpha} and \ref{eq:c_MAH_Einasto} work for a large mass range $[10^{10},2\times 10^{13}] M_{\odot}$.\footnote{The concentration $c_{\mathrm{MAH}}$ and shape parameter $\alpha_{\mathrm{MAH}}^{\rm E}$ are degenerate. Choosing different values of $\alpha_{\mathrm{MAH}}^{\rm E}$, the relation between $c_{200}$ and $c_{\mathrm{MAH}}$ will be different~\cite{Ludlow:2013bd}.}

\begin{figure}
\includegraphics[width=0.45\textwidth]{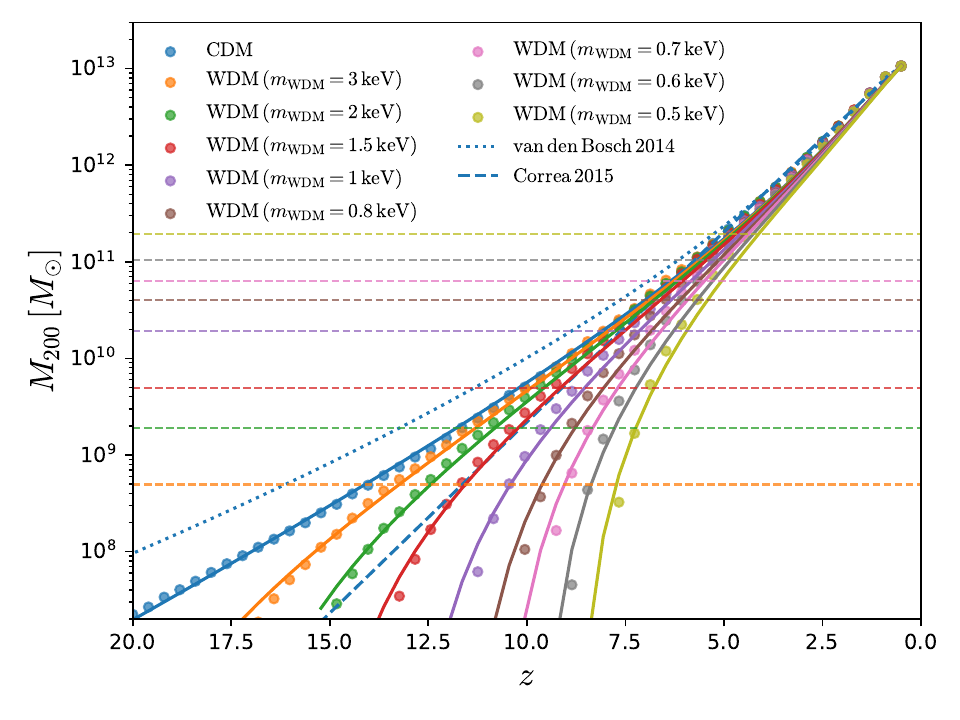}
\caption{Median mass accretion history of the hosts in WDM models compared with the result for CDM. The host mass is fixed at $10^{13} M_{\odot}$ at $z_0=0.5$. The filled circles show the results derived from merger trees. The solid curves show the model predictions. The horizontal dashed lines show the half-mode mass. In the CDM case, the fitting formulas from van den Bosch {\it et al}. 2014~\cite{vandenBosch:2014wng} (blue dashed curve) and Correa {\it et al}. 2015~\cite{Correa:2015kia} (blue dotted curve) are also shown for comparisons.}
\label{fig:MAHs_WDM}
\end{figure}

\begin{figure}
\includegraphics[width=0.45\textwidth]{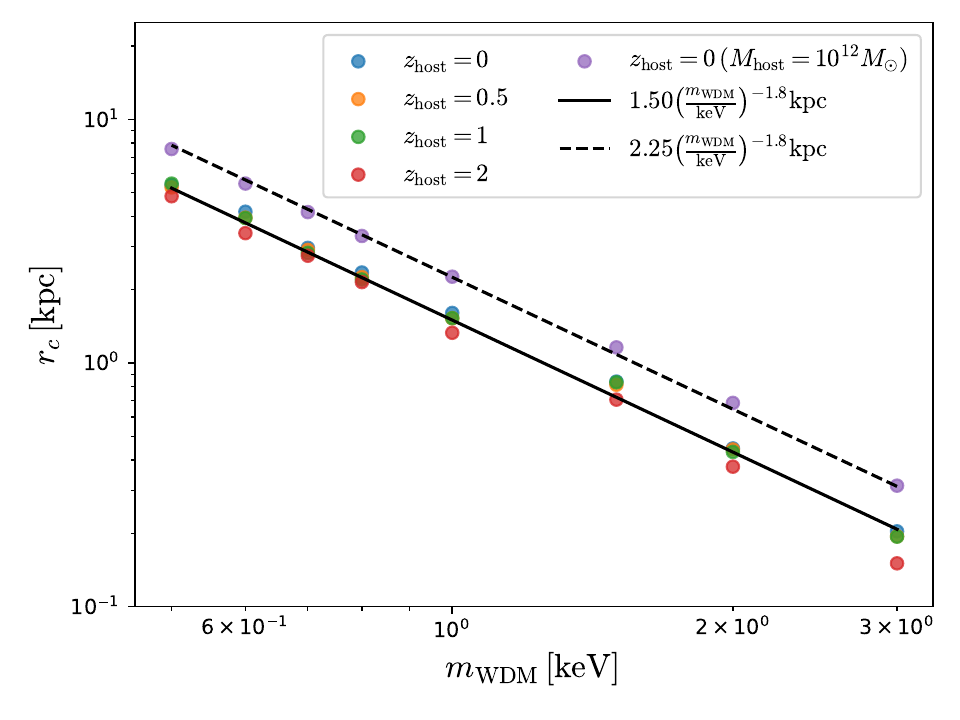}
\caption{The best-fit core radius for the MAHs in WDM models.}
\label{fig:rc_m_WDM}
\end{figure}

For dark matter models with a suppressed linear matter power spectrum on small scales compared to CDM, the MAH of the host will be different at high redshifts when the host mass is smaller. To quantify this, we have run {\sc galacticus} models using the same settings as for the CDM, but with a warm dark matter (WDM) linear matter power spectrum~\cite{Viel:2005qj}. The median MAHs for different WDM models from merger trees are shown in Fig.~\ref{fig:MAHs_WDM} (filled circles). As we can see, when the host mass is below the half-mode mass of the corresponding model (horizontal dashed lines), it begins to deviate significantly from the CDM model. This is expected because halos form later below the half-node scale. To model the mass accretion for WDM models, we consider a cored-Einasto profile~\cite{Lazar:2020pjs}
\begin{equation}
\rho_{\mathrm{Einasto}} = \rho_{-2} \exp\left[-\frac{2}{\alpha}\left(\left(\frac{r+r_{\rm c}}{r_{-2}}\right)^{\alpha}-1\right)\right],
\label{eq:cored_Einasto}
\end{equation}
where we have an addition parameter $r_{\rm c}$, the cored radius. Fixing the concentration parameter $c_{\rm MAH}$ and $\alpha_{MAH}^{\rm E}$ at the values for the CDM model, we fit the host mass growth assuming Eq.~\eqref{eq:cored_Einasto} by varying $r_{\rm c}$. We find a tight relation between $r_{\rm c}$ and the WDM particle mass,
\begin{equation}
r_{\rm c}=1.5\left(\frac{m_{\mathrm{WDM}}}{\mathrm{keV}}\right)^{-1.8}\,{\mathrm{kpc}}.
\label{eq:rc_m_WDM}
\end{equation}
The above relation is independent of the final redshift of the host (see Fig.~\ref{fig:rc_m_WDM}), but as the final host mass decreases, $r_{\rm c}$ becomes larger (dashed curve).

Again, we must emphasize that here we use the cored-Einasto profile to fit the MAH. The parameters obtained here are related to the parameters of the halo density profile, but the values are typically different. In reality, we would expect the core radius of the halo profile in WDM models to be orders of magnitude smaller~\cite{Maccio:2012qf}. For ${\mathrm{keV}}$ WDM particles, the core in the host halo is too small to affect the tidal evolution of subhalos, thus we can still approximate the host density as an NFW profile.

Because of the different host growth, subhalo evolution in the host potential will be affected. However, for realistic WDM models, e.g., $m_{\rm WDM} \gtrsim 1 {\mathrm{keV}}$, such difference is only significant at very high redshifts, $z\gtrsim 8$. Since most surviving subhalos tend to infall at lower redshift, we expect a marginal effect due to the difference in the host growth. We have taken the same population of subhalos and let them evolve in CDM and WDM hosts. We do not find a noticeable difference in the subhalo bound mass function. For more details, see Appendix~\ref{sec:SHMF_WDM}. Therefore, in the following sections, we will only consider the host growth models for CDM. We must note that, although the difference in the host growth between WDM and CDM does not affect the subhalo mass function, we do expect different subhalo mass function in WDM models at lower masses, which are captured by the suppressed infall mass function and halo concentration below the half-mode mass scale (see Sec.~\ref{ssec:shmfevolved}).\footnote{The halo concentration can be affected even at masses 1 order of magnitude higher than the half-mode scale}

\begin{figure*}
\includegraphics[width=0.45\textwidth]{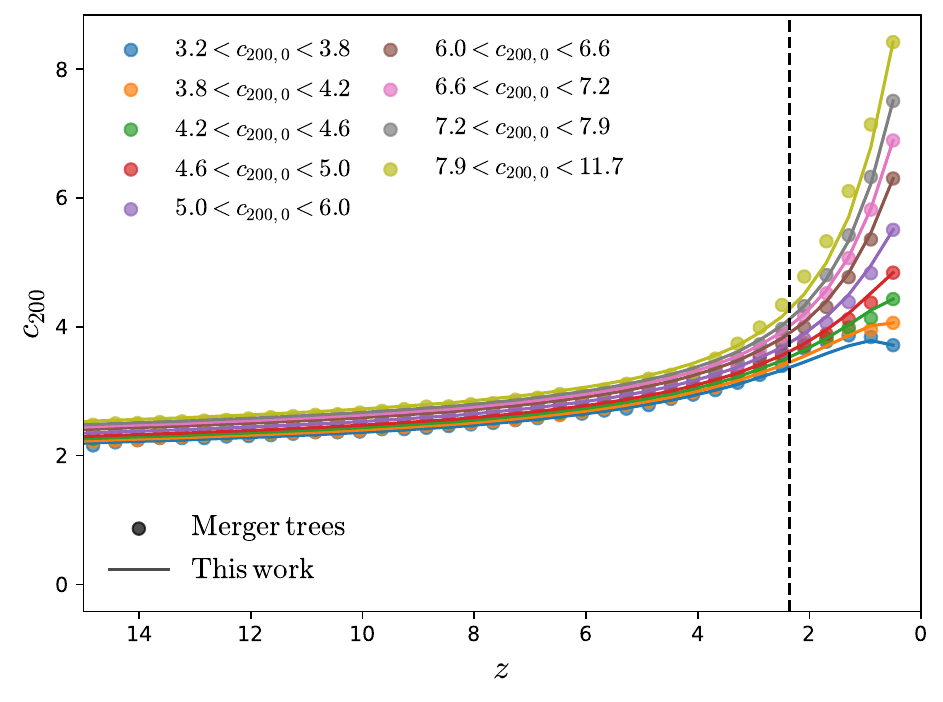}
\includegraphics[width=0.45\textwidth]{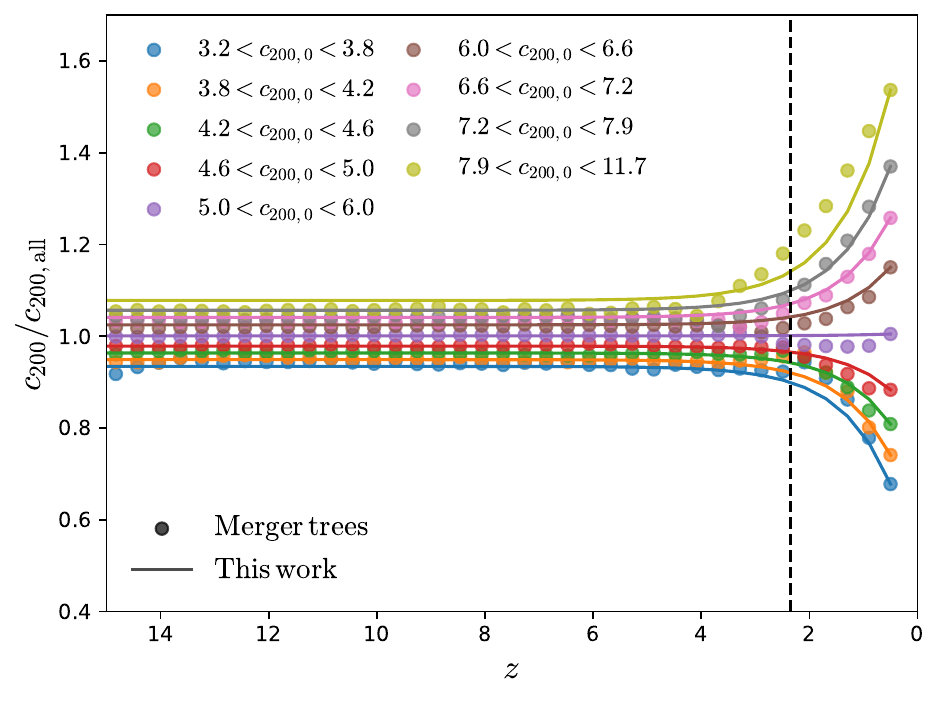}
\caption{Evolution of the host concentration binned by the current values. Filled circles are from the merger trees while the solid curves show the model proposed in this work.}
\label{fig:c200}
\end{figure*}

Apart from the host mass, we also need the host concentration to determine the host potential at redshift $z$. One simple way to do this is to utilize the median halo mass-concentration relation found in previous literature, e.g.,~\cite{Ludlow16}. However, this simple approach does not account for possible correlation between the host concentration in the past and that at the current time. As is shown in Fig.~\ref{fig:c200}, the host with a higher (lower) concentration at the current time also tends to be more concentrated in the past. The greatest differences occur after the formation redshift $z_{\rm f}$ indicated by the vertical dashed line. We propose the following formula for the host concentration:
\begin{eqnarray}
\!\!\!\!\!\!\!\!\!\!\!\! c_{200}(z) &=& c_{200}^{\mathrm{median}}(z)\left[\frac{c_{200}(z_0)}{c_{200}^{\mathrm{median}}(z_0)}\right]^{\alpha} \nonumber \\
&& \left[1+\left(\left(\frac{c_{200}(z_0)}{c_{200}^{\mathrm{median}}(z_0)}\right)^{1-\alpha}-1\right) e^{-2\frac{z-z_0}{z_{\rm f}-z_0}}\right].
\label{eq:c_host}
\end{eqnarray}
Here $c_{200}^{\mathrm{median}}(z)$ is the median concentration of hosts at redshift $z$, which can be determined by a specified halo mass-concentration relation, $\alpha=0.175$ is a parameter controlling how significantly the halo concentration deviates from the median value at $z<z_{\rm f}$. The formation redshift is defined based on the median MAH---it is that redshift at which the median host mass equals the median mass enclosed within the scale radius of the host at the current time. As can be seen from Fig.~\ref{fig:c200}, this fitting function (solid curves) can accurately describe the concentration evolution of the hosts in different concentration bins.

\section{Infall redshift distribution of subhalos}
\label{sec:infall}

Having established models for the tidal evolution of a single subhalo, to predict the subhalo statistics at the current time, one still needs to know how many subhalos fell into the host, when the infalls happened, and the properties of the subhalos at infall. In principle, we can measure this information from the merger trees extracted from cosmological simulations or synthetic trees generated using the EPS formulism. We choose the latter since it can quickly generate a large number of merger trees.

More specifically, we make use of the {\sc galacticus} code as discussed in the previous section. We have run {\sc galacticus} models with the full physics for the tidal evolution. The root halo (host) has a mass of $10^{13}M_{\odot}$ at $z_0=0.5$. We generate in total $520$ merger trees with a resolution of $M_{\rm res}=5\times 10^7 M_{\odot}$. For subhalos, we have imposed two destruction criterion~\cite{Yang:2020aqk}: (1) the subhalo is within $10^{-3} R_{\rm vir,host}$; (2) the bound mass of the subhalo drops below $5\times 10^3 M_{\odot}$. The concentration of each halo is determined from the merger history following Refs.~\cite{Ludlow16,2019MNRAS.485.5010B}.

However, as is well known that dark matter halos form hierarchically, there may be multiple levels of substructure. For example, a halo---call it halo A---merges with another (halo B); halo A is now a subhalo of B. At a later time, halo B accretes onto the main host halo; halo B will become a subhalo of C and A is now a sub-subhalo. In CDM, hierarchical structure formation iterates this process over many decades in halo mass~\cite{Diemand:2008in,Springel:2008cc,Jiang:2014nsa,Jiang:2025log}. As the sub-subhalo evolves together with the subhalo in the main host, it can finally be stripped from the subhalo and itself is promoted to become a subhalo of the main host. Based on how subhalos fell into the main host, we divide them into two groups:
\begin{itemize}
\item direct infallers which are subhalos that directly fell into the main host;
\item indirect infallers which used to be subhalos of halos other than the main host.
\end{itemize}
These two groups of subhalos may have very different tidal evolution histories.

For direct infallers, before they fell into the main host, their properties such as the halo concentration follow the relation for isolated halos. After infall, they will experience tidal effects and start to lose mass. They are well described by the models described in the previous section.

For indirect infallers, before they fell into the main host, they were subhalos of other halos and may have already lost part of their mass, and their halo concentration is defined at the time when they first became a subhalo. After infall, they will continue to lose mass. These indirect infallers are more complicated to model because we need to account for their mass loss both before and after their infall. For simplicity, we will treat these indirect infallers as direct ones but shift their infall time to an earlier time to account for possible preinfall mass loss. To this end, we define an effective infall redshift,
\begin{equation}
z_{\rm infall, eff} = w z_{\rm infall, first}+(1-w) z_{\rm infall, recent}.
\label{eq:z_infall_eff}
\end{equation}
Here $w$ is a weight factor, $z_{\rm infall,first}$ is the redshift at which a halo first becomes a subhalo (not necessarily a subhalo of the main host), and $z_{\rm infall,recent}$ is the redshift at which the halo fell into the main host. For direct infallers, $z_{\rm infall,first}=z_{\rm infall,recent}$ by definition. For indirect infallers, $z_{\rm infall,first} > z_{\rm infall,recent}$. The weight factor $w$ is adjusted to match the bound mass distribution of subhalos.

\begin{figure*}
\includegraphics[width=0.45\textwidth]{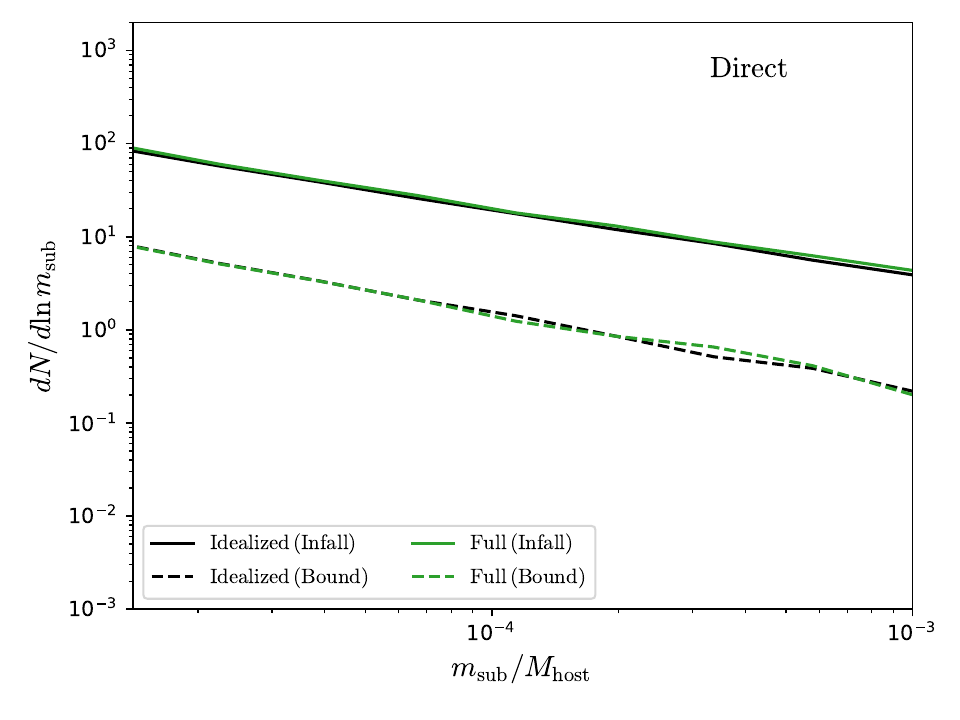}
\includegraphics[width=0.45\textwidth]{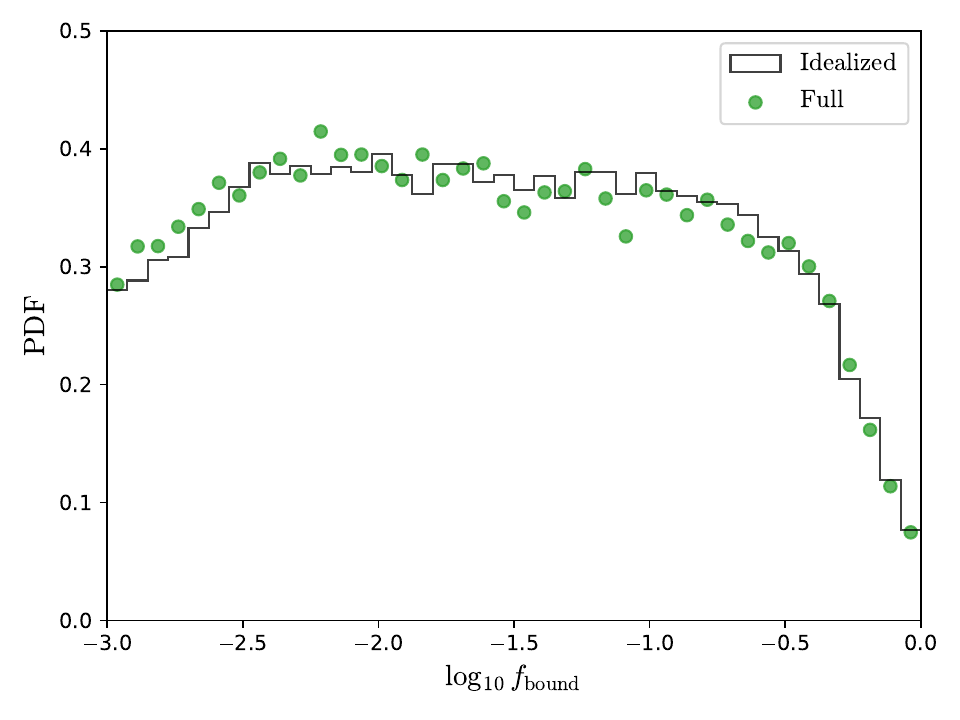}
\caption{Statistics of directly infalling subhalos from idealized runs compared with the full {\sc galacticus} models. Left panel: infall and bound mass function. Right panel: distribution function of bound mass fraction $f_{\rm b}=m_{\mathrm{bound}}/m_{\mathrm{infall}}$.}
\label{fig:direct}
\end{figure*}

\begin{figure*}
\includegraphics[width=0.45\textwidth]{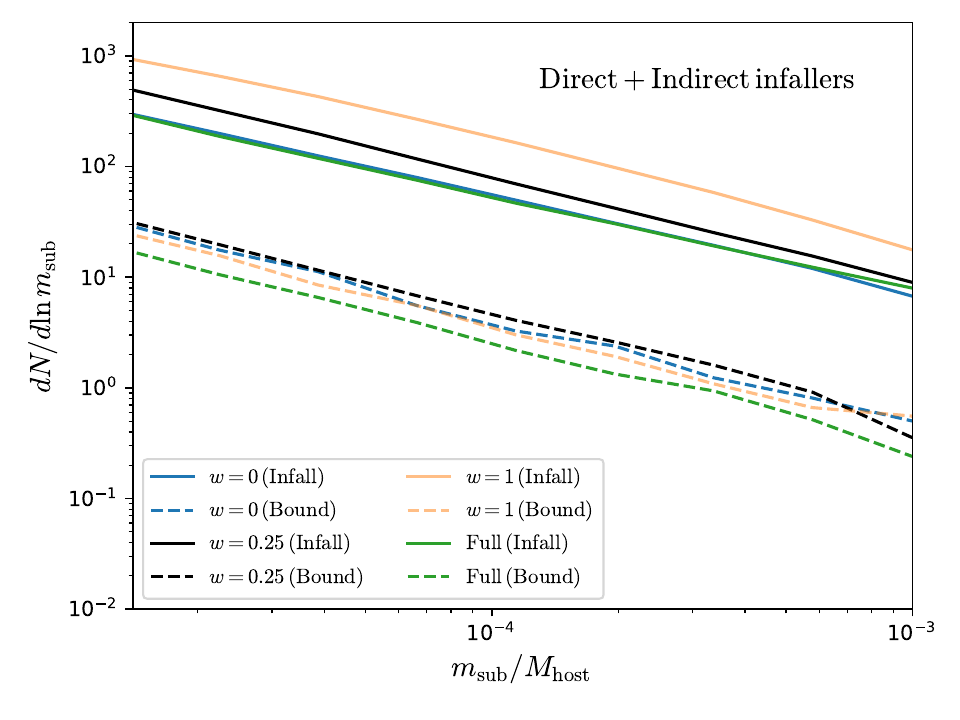}
\includegraphics[width=0.45\textwidth]{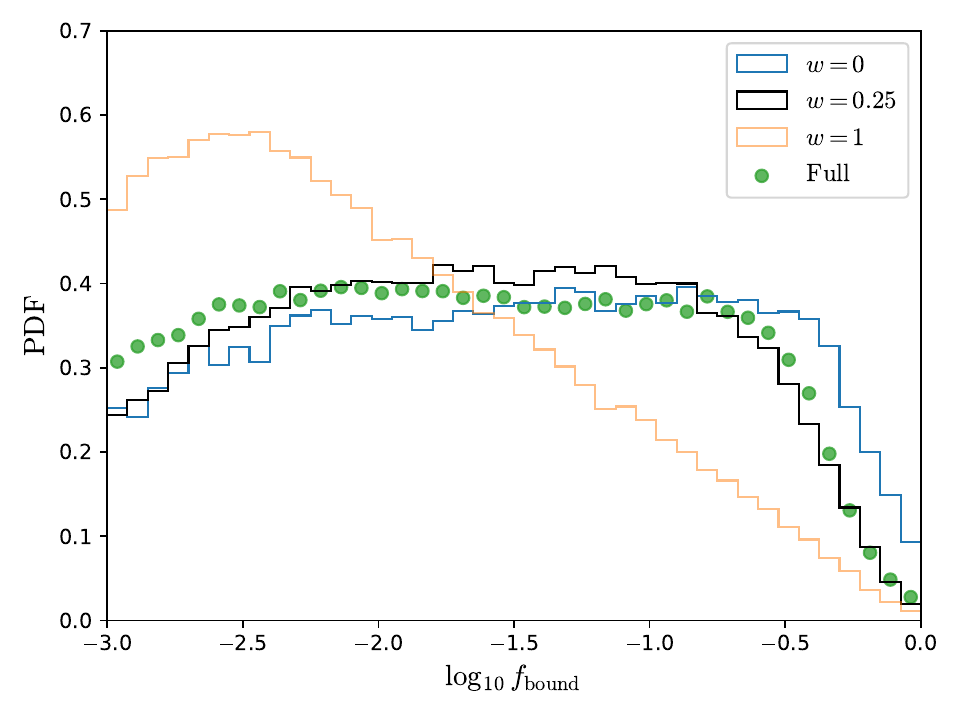}
\caption{As Fig.~\ref{fig:direct}, but including also indirect infallers. For the idealized runs, the effective infall redshift is defined as $z_{\mathrm{infall,eff}}=w z_{\mathrm{infall,first}}+(1-w) z_{\mathrm{infall,recent}}$.}
\label{fig:indirect}
\end{figure*}

To obtain the best-fit $w$, we run idealized simulations with {\sc galacticus}. This is analogous to idealized N-body simulations~\cite{Ogiya:2019del}, i.e. for each simulation there is only one single subhalo evolving in the host potential. We have used the models for the host growth in the previous section to manually create the merger trees instead of generating them using the EPS method. For each tree, we keep only the main branch and the node that contains the subhalo. The infall redshift and mass of the subhalo is drawn from the distribution functions that are measured from the full {\sc galacticus} models.

To validate our idealized simulations, we first consider only the direct infallers. We have run $10^6$ idealized simulations. Figure~\ref{fig:direct} shows statistics of these subhalos. For the purpose of lensing analysis, we focus on the central region of the host halo. The left panel shows the mass function of subhalos with a projected distance of $D<30\,{\rm kpc}$ from the center of the host in the idealized runs (black curves) compared with the mean mass functions from the full {\sc galacticus} models. Here we have normalized the results based on the infall mass function of all subhalos within the virial radius of the host. As can be seen, the bound mass function (black dashed curve) is in good agreement with that from the full model. The right panel shows the distribution function of the bound mass fraction $m_{\rm bound}/m_{\rm infall}$. Again, we see a good agreement between the idealized runs and the full models.

We then include also the indirect infallers. We find that if we take $w=1$, i.e., setting the infall redshift at $z_{\rm infall,first}$, the mass loss of subhalos is significantly overestimated, see the right panel of Fig.~\ref{fig:indirect} (orange curve). This is due to two effects. First, the indirect infaller is treated as falling into the main halo much earlier than it actually did, thus experiencing more tidal stripping.\footnote{The main host is usually more massive than the actual host of the indirect infaller at $z_{\rm infall,first}$. Consequently, the tidal stripping is stronger than it should be.} Second, the main host is smaller at earlier times, such that when the indirect infaller is added, it is closer to the center of the host. The latter effect has two impacts on the subhalo bound mass function. On the one hand, it leads to more tidal mass loss. On the other hand, it results in more subhalos within the $30\,{\rm kpc}$ projected distance. This explains why the subhalo mass function (orange curves in the left panel of Fig.~\ref{fig:indirect}) is higher than that from the full model (green curves).

For $w=0$, i.e., setting the infall redshift at the actual redshift when the subhalos entered the host's virial radius, the agreement is improved. See Fig.~\ref{fig:indirect} (blue curves). In particular,  the indirect infallers are now added to the host potential at roughly correct positions at infall. The measured infall mass function of subhalos with $D<30\,{\rm kpc}$ matches closely with that from the full models. But since the preinfall mass loss is not taken into account, the mass loss is underestimated (right panel), leading to higher subhalo bound mass function (left panel).

We find that a value of $w=0.25$ gives the best fit to the distribution of bound mass faction (see black curve in the right panel of Fig.~\ref{fig:indirect}). Similar to the case with $w=1$, the indirect infallers are closer to the host center at infall, which makes both the infall and bound mass function $\sim 40\%$ higher than that from the full model. However, we can easily rescale the infall mass function by a factor of $0.7$ to make the bound mass function match the full model.

\begin{figure}
\includegraphics[width=0.45\textwidth]{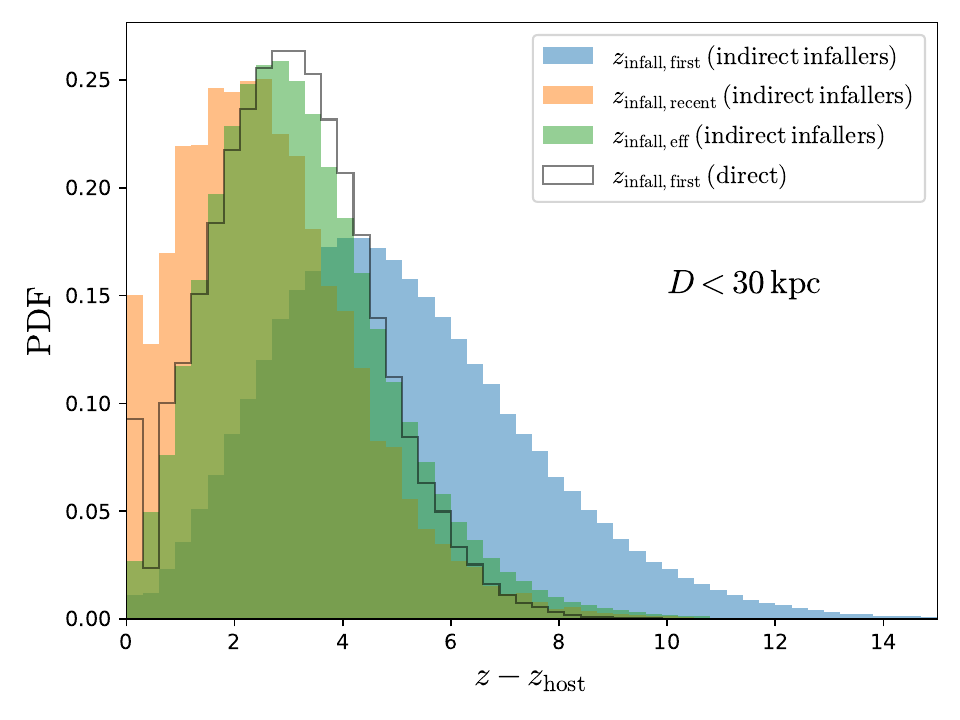}
\caption{Infall redshift distributions for direct and indirect infallers under different definitions. For direct infallers, different definitions are identical.}
\label{fig:z_infall_dist}
\end{figure}

Although the distributions of $z_{\rm infall,first}$ and $z_{\rm first,recent}$ for indirect infallers differ from those of direct infallers, we find that the distribution of $z_{\rm infall, eff}$ with $w=0.25$ is similar between the two populations when selecting subhalos with a projected distance $D < 30\,{\rm kpc}$ (see Fig.~\ref{fig:z_infall_dist}). This coincidence is worth further study in future work. For subhalos within the $30\,{\rm kpc}$ projected distance, the distribution function of $z_{\rm infall, eff}$ ($w=0.25$) can be modeled as a truncated Gaussian distribution
\begin{equation}
p(z_{\rm infall,eff}|M)=
\begin{cases}
\mathcal{A}\exp\left[-\frac{(\Delta z-\mu_z)^2}{2 \sigma_z^2}\right], & \text{if $\Delta z >0$}. \\
0, & \text{otherwise}.
\end{cases}
\label{eq:z_infall_dist}
\end{equation}
Here $\Delta_z=z_{\rm infall,eff}-z_0$, and $\mathcal{A}$ is a normalization factor. The mean $\mu_z$ and variance $\sigma_z^2$ are both mass dependent
\begin{equation}
\mu_z,\sigma_z = \frac{a}{1+b\left[\log_{10}(m_{\mathrm{sub}}/M_{\rm host})\right]^c}.
\label{eq:mu_sig_z}
\end{equation}
For $\mu_z$, the best-fit parameters are $\{a=3.35550, b=3.20547, c=-2.91076\}$. For $\sigma_z$, the best-fit parameters are $\{a =1.97880, b=4.17391, c=-2.14428\}$.

\begin{figure}
\includegraphics[width=0.45\textwidth]{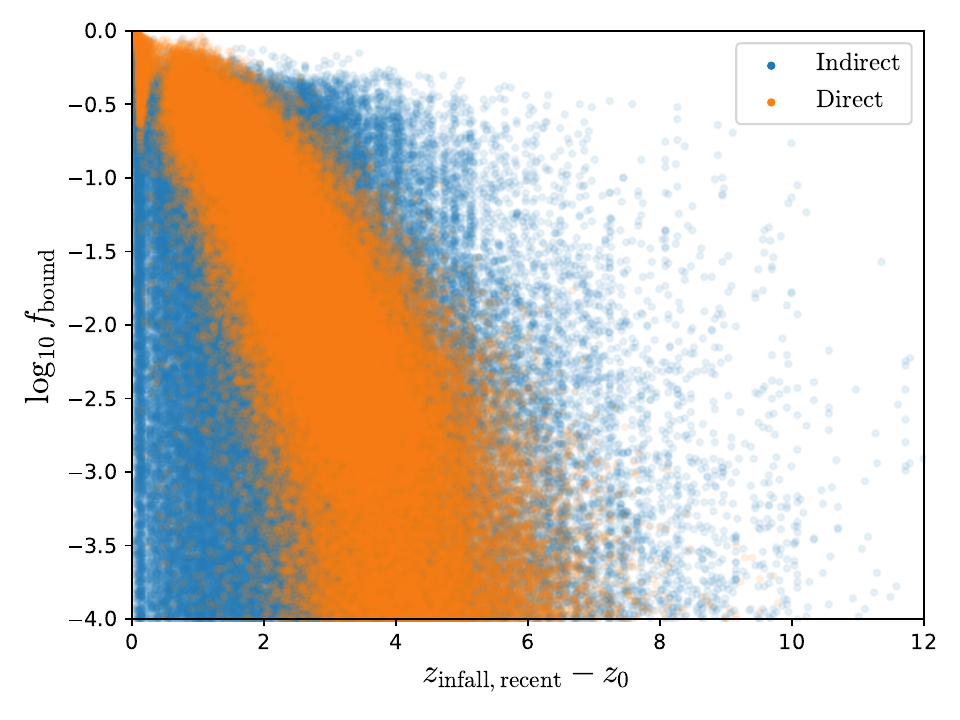}
\caption{Bound mass fraction of subhalos with respect to $z_{\rm infall, recent}$ from the full {\sc galacticus} model.}
\label{fig:f_b_z_infall}
\end{figure}

To further demonstrate the difference between the direct and indirect infallers, we show the bound mass fraction with respect to the redshift when the subhalo entered the host's virial radius in Fig.~\ref{fig:f_b_z_infall}. Each point represents one subhalo from the full {\sc galacticus} model. As can be seen, for the same $z_{\rm infall, recent}$, the bound mass fraction of indirect infallers shows a much larger scatter. This is due to two reasons: (1) before infall onto the main halo, these subhalos may already lose some mass, making the mass fraction lower than for a direct infaller; (2) the indirect infaller tends to form earlier when the universe is denser, so they are compact and less affected by tidal effects. Consequently, some of them have a higher bound mass fraction than the direct infaller. By defining an effective infall redshift as Eq.~\eqref{eq:z_infall_eff}, we can match the median mass loss of indirect infallers as shown in Fig.~\ref{fig:indirect}, but will underestimate the scatter in this quantity. We will come back to this point later in Sec.~\ref{ssec:singlehalosims}.

\section{A model for subhalo tidal evolution around strong lenses}
\label{sec:pyhalosims}

In this section, we present a model for the tidal evolution of dark substructure around a typical strong lens system. The model is intended to rapidly (in $\sim$ seconds) generate populations of dark subhalos that appear in projection near the Einstein radius of a typical lens system in such a way that we reproduce the bound mass function and subhalo density profiles predicted by N-body simulations and {\sc galacticus} models. This model is implemented in the open-source software {\tt{pyHalo}}\footnote{https://github.com/dangilman/pyHalo} for use in forthcoming strong lensing inferences of dark matter substructure. 

In Sec.~\ref{ssec:singlehalosims}, we begin by describing the calibration of this tidal stripping model, making use of idealized {\sc galacticus} simulations (in which a single halo evolves over cosmic time in a growing host potential, as discussed in Sec.~\ref{ssec:hostevolution}) to predict the bound mass function at $z=0.5$ in a group-scale $\sim 10^{13} M_{\odot}$ host. In Sec.~\ref{ssec:shmfevolved} we compare the predictions of our model with the full {\sc galacticus} model. Throughout this section, when we discuss an infall redshift, we refer to the ``effective" infall redshift, as discussed in Sec.~\ref{sec:infall}. 

\subsection{Predicting the subhalo bound mass function}
\label{ssec:singlehalosims}

We use the idealized {\sc galacticus} model discussed in the previous section to compute $p\left(f_{\rm{bound}}| z_{\rm{infall}}, c_{\rm{sub}}, c_{\rm{host}},r_{\rm{2D}}<30\,\rm{kpc}\right)$, the probability that a subhalo has a bound mass of $m_{\rm{bound}}$ at $z = 0.5$, given that it had a concentration at infall $c_{\rm{sub}}$, it was accreted at redshift $z = z_{\rm{infall}}$, the host has $M_{200} = 10^{13} M_{\odot}$ and concentration $c_{\rm{host}}$ at $z=0.5$, and that the subhalo appears in projection within $30 \ \rm{kpc}$ of the host center. These conditions are appropriate for a typical strong lens system with an Einstein radius of $\sim 1 $ arcsec, which corresponds to a projected radius of $6$--$30$ kpc. We emphasize that, although we fix the host redshift in our analysis, the model presented below is also applicable to hosts at other redshifts. This is because we have written the infall redshift distribution of subhalos in terms of $z_{\rm infall}-z_{\rm host}$ and the host's mass growth also depends primarily on $z-z_{\rm host}$~\cite{Correa:2015dva}. Besides, more massive hosts tend to contain more subhalos. Such dependence can be absorbed into the normalization parameter of the infall mass function provided that the host mass does not differ significantly from the one considered in this work. For a more comprehensive discussion on how the normalization of the subhalo mass function depends on the host mass and redshift, we refer readers to~\cite{Gannon:2025nhr}.

To determine the distribution $p\left(f_{\rm{bound}}| z_{\rm{infall}}, c_{\rm{sub}}, c_{\rm{host}},r_{\rm{2D}}<30\,\rm{kpc}\right)$, we use the idealized {\sc galacticus} model to evolve subhalos from infall until $z=0.5$ with various combinations of $\left(z_{\rm{infall}}, c_{\rm{sub}}, c_{\rm{host}}\right)$. We perform these calculations on a tabulated grid of $z_{\rm{infall}} \in \left[0.75, 8\right]$, $\log_{10} c_{\rm{sub}} \in \left[0.3, 2.5\right]$, and $c_{\rm{host}} \in \left[3,9\right]$. For each coordinate on the tabulated grid we use {\sc galacticus} to generate initial orbital trajectories, and track the subhalos from infall at $z = z_{\rm{infall}}$ until $z = 0.5$ in an evolving host potential, as described in Sec.~\ref{sec:infall}. If the subhalo appears in projection within $30 \ \rm{kpc}$ of the host halo center at the end of the simulation we record its final bound mass $m_{\rm{bound}}$. 

Following these calculations, for each combination of $\left(z_{\rm{infall}}, c_{\rm{sub}}, c_{\rm{host}}\right)$ we have a distribution of $f_{\rm{bound}}=m_{\rm{bound}}/m_{\rm{infall}}$. We have verified that the tidal stripping has a negligible mass dependence, so we can omit an explicit dependence on the subhalo's infall mass in these calculations. For a fixed set of parameters $c_{\rm{sub}}$, $z_{\rm{infall}}$, $c_{\rm{host}}$, the scatter in $f_{\rm{bound}}$ comes from different subhalo orbits: some objects pass close to the host center and lose large amounts of material, while others maintain a large physical separation from the host center, and thereby avoid severe tidal stripping, but appear in projection near the Einstein.

\begin{figure*}
	\includegraphics[trim=0.3cm 1cm 0.25cm
	0.5cm,width=0.32\textwidth]{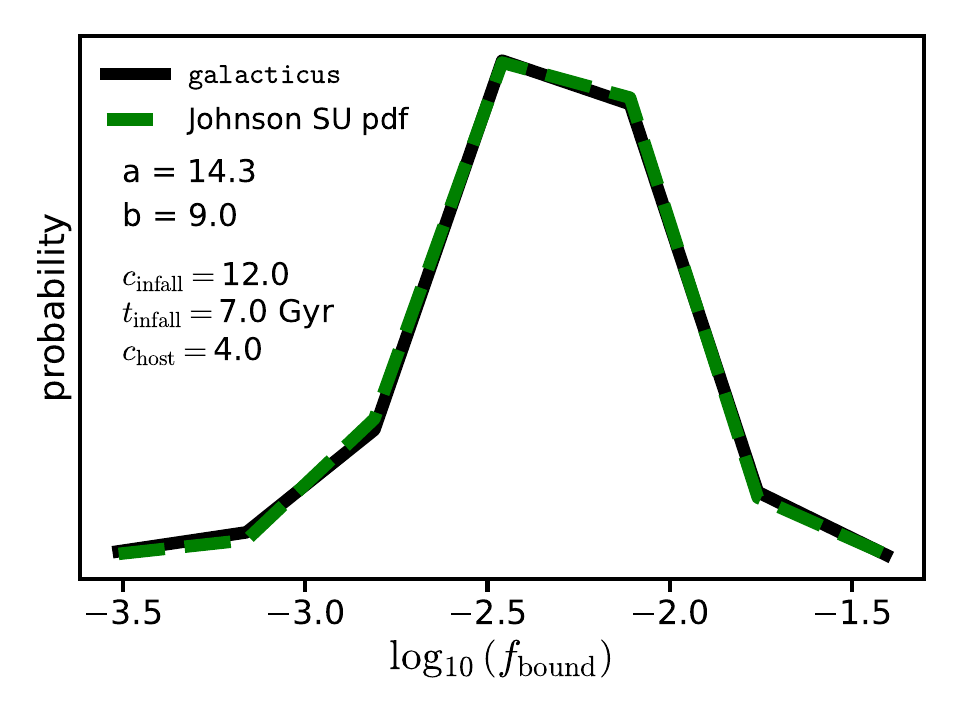}
 \includegraphics[trim=0.3cm 1cm 0.25cm
	0.5cm,width=0.32\textwidth]{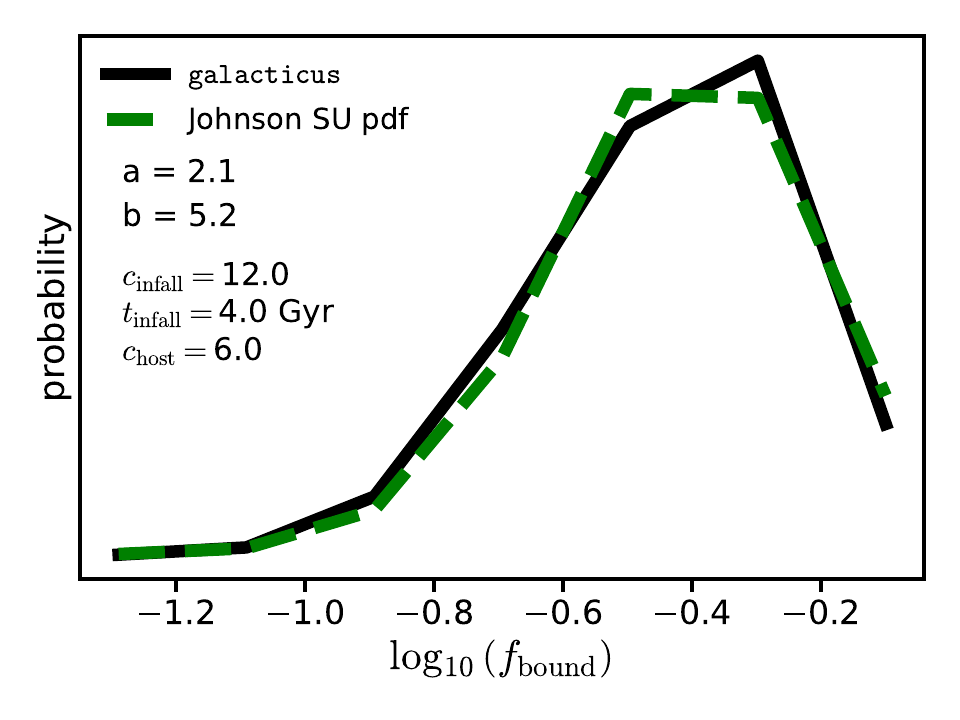}
 \includegraphics[trim=0.3cm 1cm 0.25cm
	0.5cm,width=0.32\textwidth]{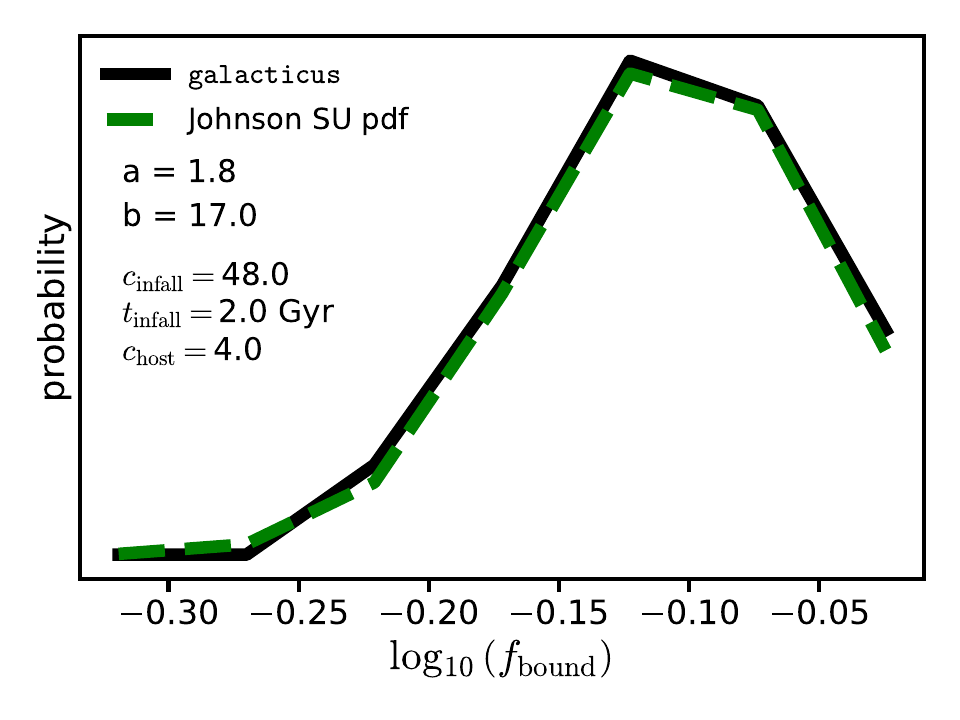}
	\caption{\label{fig:pdffits} The distribution of bound masses predicted by idealized {\sc galacticus} (black curve) simulations in which we inject single subhalos into an evolving host (see discussion in Sec.~\ref{ssec:singlehalosims}). We fit these distributions with Johnson SU probability density functions with parameters $a$ and $b$ [Eq.~\eqref{eqn:johnsonpdf}, green dashed curve]. Each panel shows a different combination of subhalo infall concentration, elapsed time since infall, and host halo concentration at $z=0.5$. Using spline interpolations of $a$ and $b$, we obtain a continuous, probabilistic model for the bound mass fraction $f_{\rm{bound}}$ (see Fig.~\ref{fig:curves4}).}
\end{figure*}

The solid black curves in Fig.~\ref{fig:pdffits} show three examples of the distribution of $f_{\rm{bound}} = m_{\rm{bound}}/m_{\rm{infall}}$ for various combinations of $z_{\rm{infall}}, c_{\rm{sub}}, c_{\rm{host}}$. For easier interpretation, we have replaced $z_{\rm{infall}}$ with the time since infall, $t_{\rm{infall}}$, in each panel. The dashed red curve is a Johnson SU distribution~\cite{JohnsonSU} for $x\equiv \log_{10} f_{\rm{bound}}$
\begin{equation}
\label{eqn:johnsonpdf}
p\left(x\right) = \frac{b}{\sqrt{2 \pi \left(x^2+1\right)}} \exp\left[-\frac{1}{2}\left(a + b \  \rm{sinh}^{-1} x\right)^2\right]
\end{equation}
with free parameters $a$ and $b$ that we have fit to match the distributions output by $\tt{galacticus}$.  

\begin{figure}
	\includegraphics[trim=0.5cm 0.5cm 0.5cm
	0.5cm,width=0.48\textwidth]{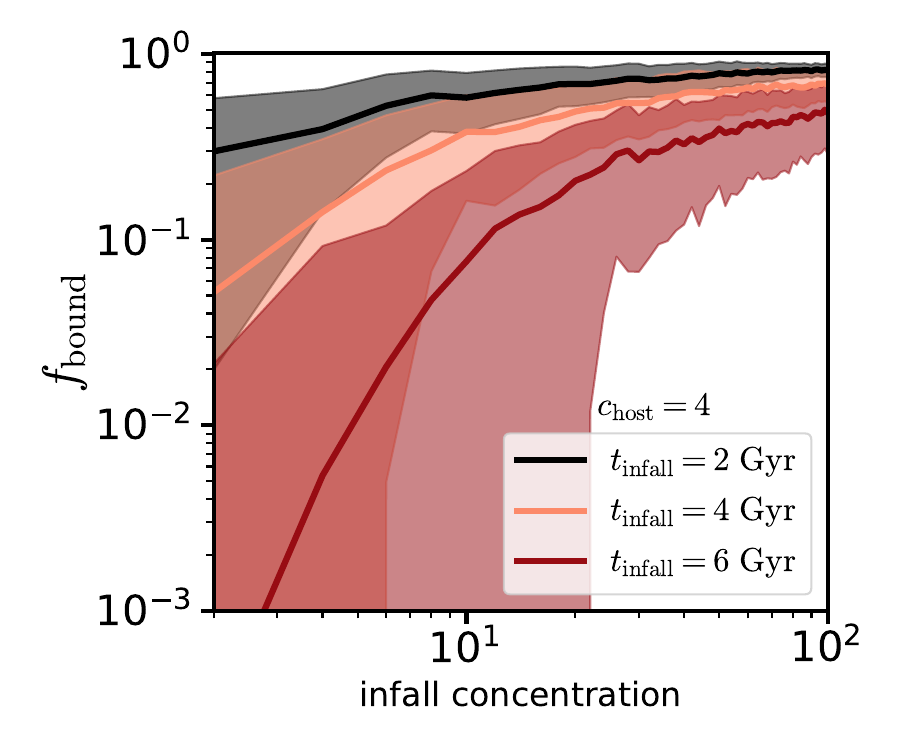}
    \caption{\label{fig:curves4} The bound mass fractions $f_{\rm{bound}}$ predicted by the tidal stripping model presented in Sec.~\ref{ssec:singlehalosims}. The curves represent the median and $68\%$ confidence intervals of the distribution of bound masses as a function of the time since infall ($t_{\rm{infall}}$), infall concentration, and $c_{\rm{host}}=4$. }
\end{figure}
\begin{figure*}
    \includegraphics[trim=1cm 0cm 0cm
	1cm,width=0.48\textwidth]{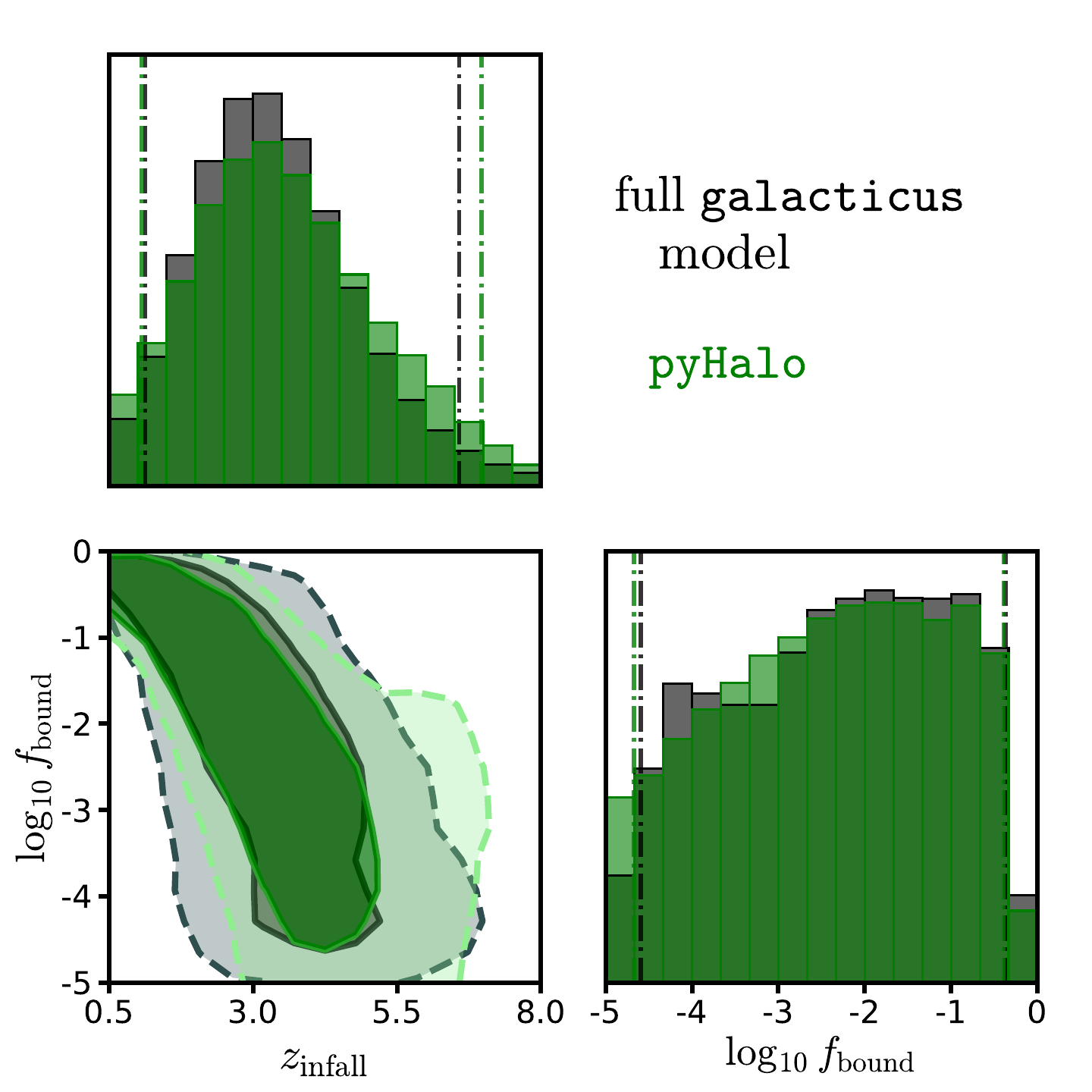}
    \includegraphics[trim=1cm 0cm 0cm
	1cm,width=0.48\textwidth]{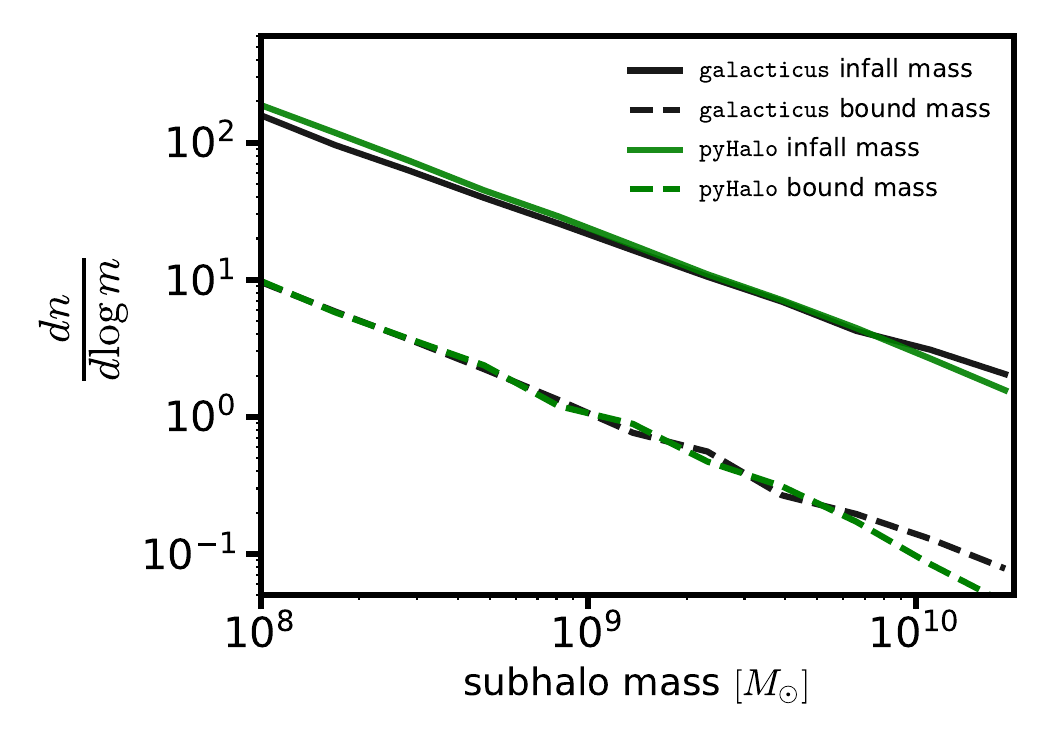}
    \caption{\label{fig:boundmassfunc} {\bf{Left:}} the joint distribution of infall redshift and the bound mass fraction $f_{\rm{bound}}$ of subhalos predicted by {\sc galacticus} (black) and {\tt{pyHalo}} (green). {\bf{Right:}} the infall (solid lines) and bound (dashed lines) mass functions predicted by {\sc galacticus} and {\tt{pyHalo}}. The tidal stripping model used in {\tt{pyHalo}} is described in Sec.~\ref{ssec:singlehalosims}. The bound mass function is suppressed by a factor of $20$.}
\end{figure*}
\begin{figure}
   \includegraphics[trim=0cm 0cm 0cm
	1cm,width=0.45\textwidth]{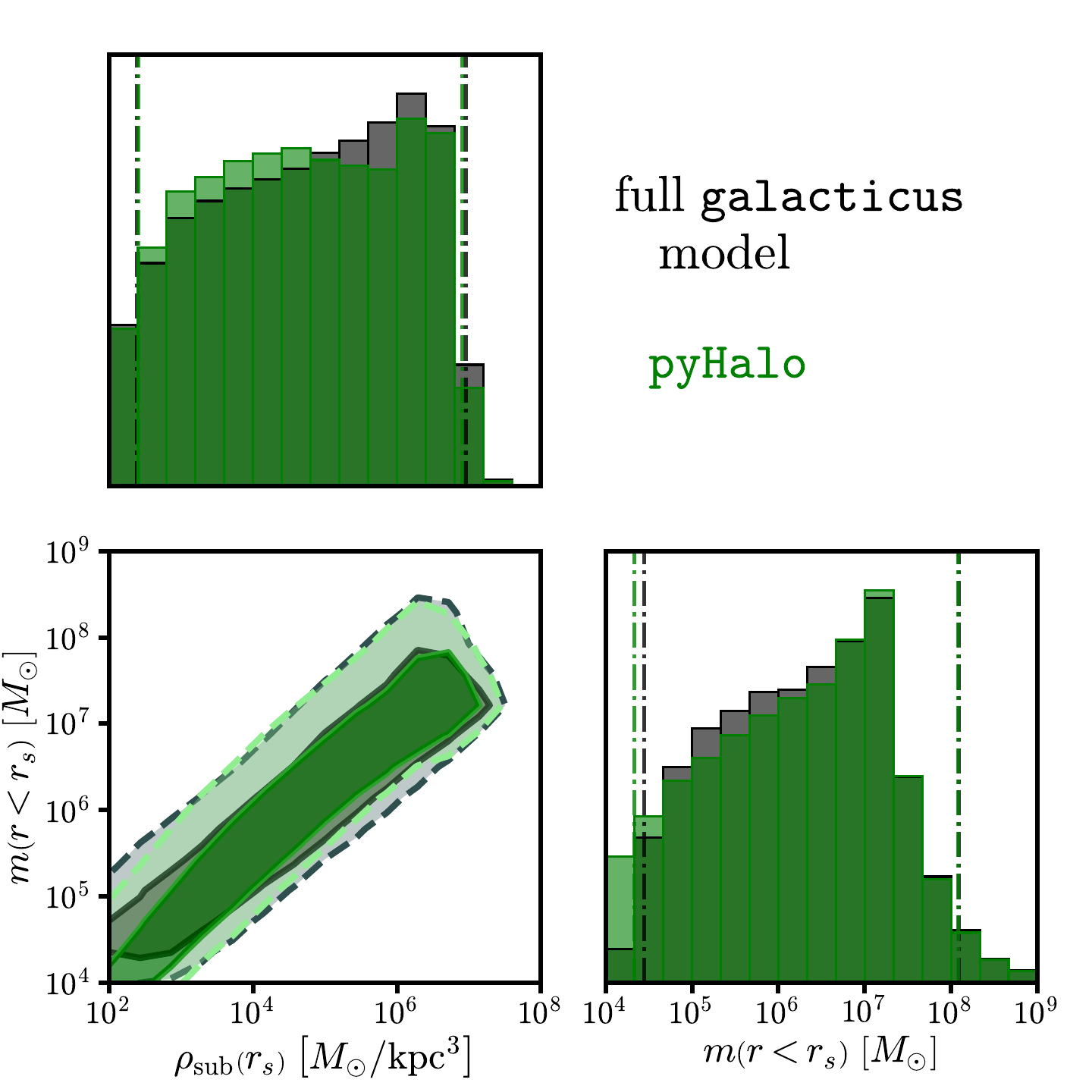}
	\caption{\label{fig:rhorsmrs} The joint distribution of the mass enclosed inside the infall scale radius, $m\left(r<r_{\rm s}\right)$, and the density profile of tidally stripped subhalos evaluated at the infall scale radius, $\rho_{\rm{sub}}\left(r_{\rm s}\right)$, between {\sc galacticus} and {\tt{pyHalo}}. Solid and dashed contours correspond to $68 \%$ and $90 \%$ confidence intervals.}
\end{figure}
Next, we perform a spline interpolation of the shape parameters $a$ and $b$ of the Johnson distribution for each combination of $z_{\rm{infall}}, c_{\rm{sub}}, c_{\rm{host}}$. The $a$ and $b$ parameters in the Johnson distribution are now functions of physical quantities, i.e., $a\left(z_{\rm{infall}}, c_{\rm{sub}}, c_{\rm{host}}\right)$, $b\left(z_{\rm{infall}}, c_{\rm{sub}}, c_{\rm{host}}\right)$. Given the infall mass, redshift, and concentration of a subhalo, as well as the host halo concentration at $z=0.5$, we can then evaluate $a$ and $b$, and draw a sample from the corresponding probability distribution in Eq.~\eqref{eqn:johnsonpdf} to obtain $m_{\rm{bound}}$. 

Given $m_{\rm{bound}}$, we use the subhalo tidal evolution model presented by \cite{Du:2024sbt} to obtain the subhalo density profile, which is the last step needed to calculate the lensing deflection angle produced by the subhalo. The density profile is assumed to follow a truncated NFW profile:
\begin{equation}
\label{eqn:rhonfw}
\rho\left(r,\rho_s,r_{\rm s},f_{\rm{bound}}\right) = \rho_{\rm{NFW}}\left(r,\rho_s,r_{\rm s}\right) T\left(f_{\rm{bound}}\right)
\end{equation}
where $\rho_{\rm{NFW}}$ represents an NFW profile, and 
\begin{equation}
T\left(f_{\rm{bound}}\right) = \frac{f_{\rm t}}{1+(r/r_{\rm t})^2}
\label{eq:rho_T}
\end{equation}
is a transfer function for the tidally stripped density profile described by \cite{Du:2024sbt}, calibrated against N-body simulations. In $T\left(f_{\rm{bound}}\right)$, both $f_{\rm t}$ and $r_{\rm t}$ are functions of $f_{\rm{bound}}$, so tidal stripping causes an overall rescaling and a truncation of the profile. Note that we do not directly use the relations between $f_{\rm t}$, $r_{\rm t}$, and $f_{\mathrm{bound}}$ found in \cite{Du:2024sbt}, because the halo mass definition is redshift dependent. Instead we use the mass ratio $m_{\mathrm{bound}}/M_{\mathrm{mx,0}}$ as a proxy. Here $M_{\mathrm{mx,0}}$ is initial mass enclosed within the maximum circular velocity radius at infall. See more details in Appendix~\ref{app:track}.

In summary, given the host halo concentration at $z=0.5$ and the concentration of a subhalo at the effective infall time, as described in Sec.~\ref{sec:infall}, we can predict the probability that a subhalo has a bound mass $m_{\rm{bound}}= f_{\rm{bound}} m_{\rm{infall}}$ and appears in projection near the Einstein radius of a typical strong lens system. This model works extremely well for predicting the median bound mass fraction of subhalos obtained from full-physics computations with $\tt{galacticus}$. However, the model calibrated following the steps outlined in this section underpredicts the scatter in $f_{\rm{bound}}$ at fixed $z_{\rm{infall}}$, and underestimates the median mass loss by a factor of $\sim 1.8$. This is due to the fact that we do not account for scatter in the host halo's MAH, and complicated tidal evolution of sub-subhalos that the idealized simulations do not fully capture, even with the definition of the infall redshift that accounts for sub-subhalo evolution. To improve the agreement between the model described in this section and the full-physics calculations, we make two small empirical modifications: first, for all $c_{\rm{host}}$, $z_{\rm{infall}}$ and $c_{\rm{infall}}$ distributions we add Gaussian intrinsic scatter to $\log f_{\rm{bound}}$ with a mean of $0.0$ and a standard deviation $|0.5 \times \log_{\rm{10}} f_{\rm{bound}}|$. Second, we add a systematic additional mass loss of $-0.25$ dex. These modifications are tuned to reproduce the joint distribution of infall redshift and the bound mass fraction obtained from the full {\sc galacticus} model. A more physically motivated model will be explored in future work.

\subsection{The evolved subhalo mass function}
\label{ssec:shmfevolved}

We now compare the predictions of the model discussed in the previous subsection with full {\sc galacticus} calculations. For these comparisons, we generate $179$ merger trees with a minimum merger tree mass limit of $5\times 10^{7} M_{\odot}$, and track subhalo evolution in the host until it stripped below $5\times 10^{3} M_{\odot}$. The tidal stripping model discussed in the previous section is implemented in {\tt{pyHalo}}. Generating the $200$ realizations of subhalo populations with {\tt{pyHalo}} we use for comparison with $\tt{galacticus}$ takes $\sim 2$ minutes on a single CPU core. 
\begin{figure}
   \includegraphics[trim=0cm 1cm 0cm
	1cm,width=0.45\textwidth]{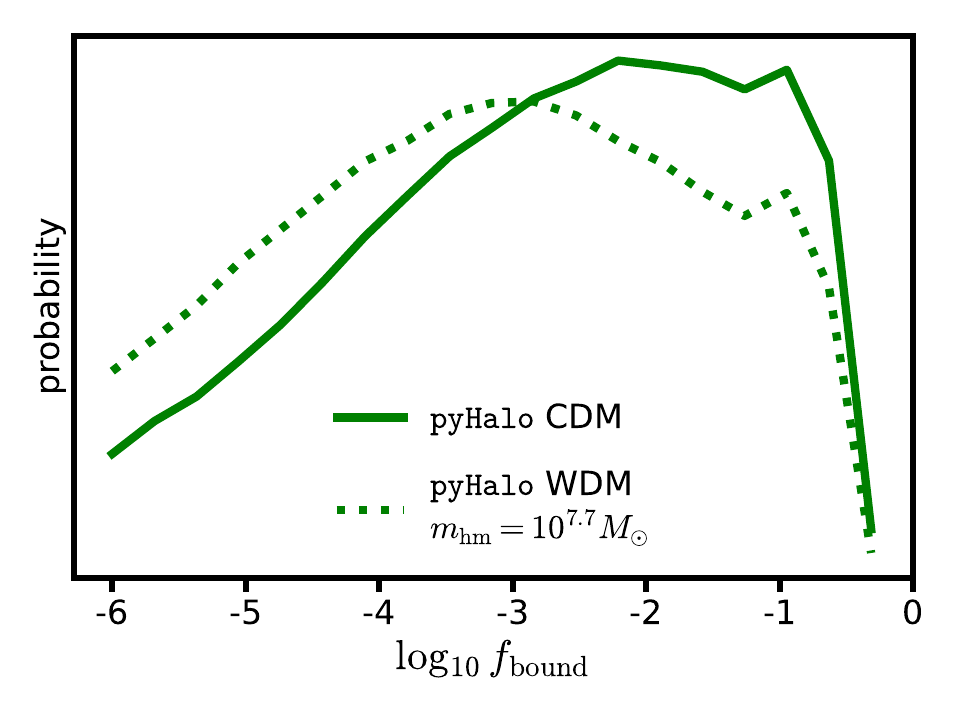}
	\caption{\label{fig:mboundwdm} The distribution of bound mass fractions in CDM (solid line) and in a WDM model (dotted) with a suppression of the halo mass function at $m \sim 5 \times 10^7 M_{\odot}$, as predicted by the tidal stripping model implemented in {\tt{pyHalo}}. The lower concentrations at infall in WDM models lead to more severe tidal stripping.}
\end{figure}

Figure \ref{fig:curves4} shows the degree of subhalo mass loss, $f_{\rm{{bound}}}$, predicted by the model discussed in Sec.~\ref{ssec:singlehalosims} as a function of the infall concentration and the elapsed time since infall. Several trends are apparent from the model predictions: halos with lower concentrations at infall are more likely to experience significant mass loss; halos that were accreted earlier tend to experience more significant mass loss; there is a large scatter in the mass loss experienced by subhalos, particularly for low infall concentrations. 

The calibration of the tidal stripping model did not rely on a concentration-mass relation, and computed the probability of $f_{\rm{bound}}$ given a subhalo's infall concentration and infall time with these variables varying independently. This allows us to use the model with any concentration-mass relation. In the calculations that follow, we use the concentration-mass relation presented by \cite{Ludlow:2016ifl}. We evaluate the concentration-mass relation at the effective infall redshift, $z_{\rm{infall,eff}}$, and compute the $\rho_s$ and $r_{\rm s}$ parameters of the NFW profile with respect to the critical density at $z_{\rm{infall,eff}}$. 

In Fig.~\ref{fig:boundmassfunc}, the left panel shows the joint distribution of the infall redshift and the bound mass fraction predicted by full {\sc galacticus} models (black), and the model presented in this work that we have implemented in {\tt{pyHalo}} (green). The right panel shows the bound and infall mass functions. {\tt{pyHalo}} uses a parametric model for the infall mass function (solid curves) 
\begin{equation}
\frac{d^2N}{d \log m d A} = \Sigma_{\rm{sub}}\left(m/m_0\right)^{-\alpha},
\end{equation}
which gives the number of subhalos per logarithmic mass interval per unit projected area. A logarithmic slope $\alpha=0.9$ and normalization $\Sigma_{\rm{sub}} = 0.11 \,\rm{kpc^{-2}}$ matches the infall mass function shown in Fig.~\ref{fig:boundmassfunc}, and application of the tidal stripping model discussed in Sec.~\ref{ssec:singlehalosims} yields the bound mass function (dashed curves). With the effective infall redshift distribution given by Eq.~\eqref{eq:z_infall_dist}, the bound mass function from {\tt{pyHalo}} agrees well with that from the {\sc galacticus} model which includes a full hierarchical treatment of subhalos. For subhalos that appear in projection within $30\,\rm{kpc}$ of the host halo center, the bound mass function is suppressed relative to the infall mass function by a factor of about $20$, on average. This corresponds to a projected surface mass density of $3.2 \pm 0.6 \times 10^6 M_{\odot}\,\rm{kpc^{-2}}$ in the bound mass range $10^8$--$10^{10} M_{\odot}$ near the Einstein radius of a typical strong lens system. 

The tidal stripping model discussed in Sec.~\ref{ssec:singlehalosims} is specifically calibrated to reproduce the bound mass function predicted by ${\sc galacticus}$ simulations, so it is unsurprising to find excellent agreement when comparing the full-physics {\sc galacticus} simulations with our model. However, the tidal stripping model we develop also provides an excellent representation of the density profiles of tidally evolved subhalos, a highly nontrivial outcome given the complexities associated with tidal evolution, and our use of a parametric form for the subhalo density profile in {\tt{pyHalo}}. In Fig.~\ref{fig:rhorsmrs}, we show the joint distribution of the subhalo density at the infall scale radius $\rho_{\rm{sub}}\left(r_{\rm s}\right)$ and the mass enclosed within the infall scale radius, $m\left(r<r_{\rm s}\right)$. The black contours represent the output from {\sc galacticus} models, which account for the fully nonlinear evolution of subhalo density profiles subject to tidal stripping, tidal heating, etc., and do not enforce a parametric form for the density profile. The green contours show the predictions of our tidal stripping model implemented in {\tt{pyHalo}}, where we model subhalo density profiles with the parametric form given in Eq.~\eqref{eqn:rhonfw} and evaluate the bound masses as described in Sec.~\ref{ssec:singlehalosims}. The agreement between these distributions indicates that we can accurately predict the internal structure of dark subhalos, because $\rho_{\rm{sub}}\left(r_{\rm s}\right)$ and $m\left(r<r_{\rm s}\right)$ constrain both the density of the subhalo at $r_{\rm s}$ and the integral of the density. 

Before using the tidal stripping implementation in {\tt{pyHalo}} to examine how this affects observables in strong lens systems, we point out that the model we develop has additional implications for tidal stripping in dark matter models with suppressed small-scale power, such as warm dark matter (WDM). In WDM scenarios, a cutoff in the matter power spectrum precludes the formation of low-mass halos, and also suppresses halo concentrations, e.g.,~\cite{Bode01,Ludlow16}. As discussed in Appendix \ref{sec:SHMF_WDM}, we can use the model for tidal stripping for WDM scenarios because the evolution of the $\sim 10^{13} M_{\odot}$ host is largely unaffected by the small-scale suppression of the power spectrum. In Fig.~\ref{fig:mboundwdm} we show the distribution of bound mass fractions predicted in CDM (solid green curves) and a WDM model with a half-mode cutoff scale $m_{\rm{hm}} = 10^{7.7} M_{\odot}$. We use the WDM concentration-mass relation presented by \cite{Ludlow16}, and the model for the suppression of the halo mass function by \cite{Lovell20}. Without a tidal stripping model that accounts for the dependence on infall concentration, these distributions would be identical. However, as our framework includes an explicit dependence on subhalo concentration at infall, our model predicts more severe tidal stripping for the less-concentrated WDM halos. This further distinguishes the lensing signals of dark subhalos in WDM from CDM. A comprehensive analysis of lensing constraints on CDM and WDM using the techniques presented in this work is currently underway~\cite{Daniel:2025}.

\section{Strong lensing signatures of tidally evolved subhalos}
\label{ssec:lensingsignatures}

\begin{figure}
	\includegraphics[trim=0.5cm 1.5cm 1cm
	2cm,width=0.45\textwidth]{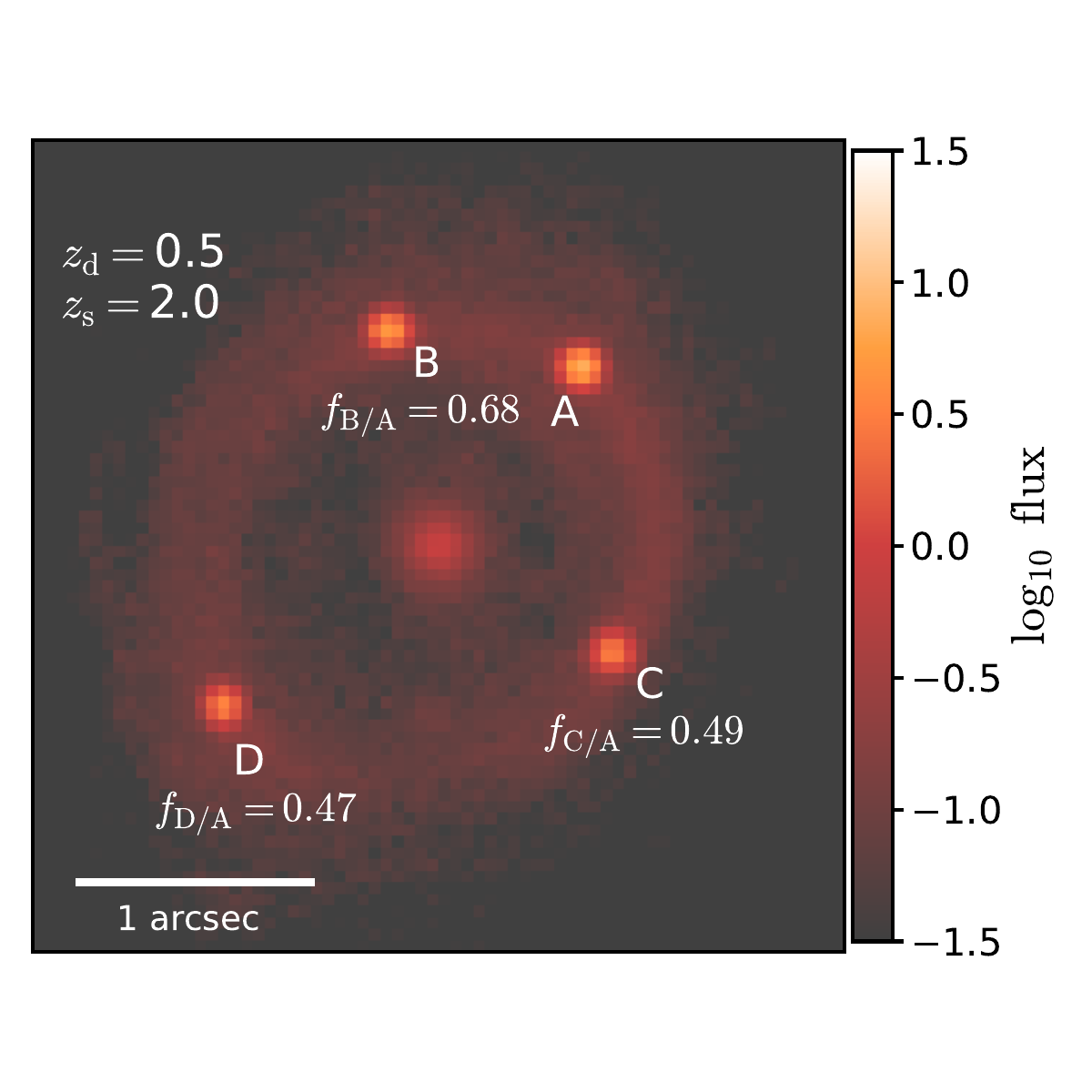}
	\caption{\label{fig:mockimage} The mock lens used to calculate the magnification cross section shown in Fig.~\ref{fig:magcross} and the flux ratio statistics shown in Fig.~\ref{fig:frcumulative}. The four image positions and flux ratios are marked in white. We use an image of a spiral galaxy extracted from the COSMOS catalog as the quasar host galaxy.}
\end{figure}

We focus on the observational implications for strongly lensed quasars, systems in which a background quasar becomes doubly or quadruply imaged by a foreground galaxy. The relative image magnifications, or flux ratios, depend on the second derivatives of the gravitational potential projected onto the plane of the lens, which makes these data particularly sensitive to the abundance and internal structure of dark-matter subhalos that appear near a lensed image in projection. In this section, we do not aim to rederive existing bounds on the nature of dark matter using our improved tidal stripping framework. Instead, our goal is to develop intuition for how our model will affect future results. To this end, we consider a mock lens system shown in Fig.~\ref{fig:mockimage}, and compute the sensitivity of the brightest image (image A) to the presence of a single tidally stripped subhalo. We will then consider the effects of a population of tidally stripped subhalos relative to a population of nontruncated NFW profiles.

\begin{figure*}
	\includegraphics[trim=1cm 1cm 0cm
	0cm,width=0.48\textwidth]{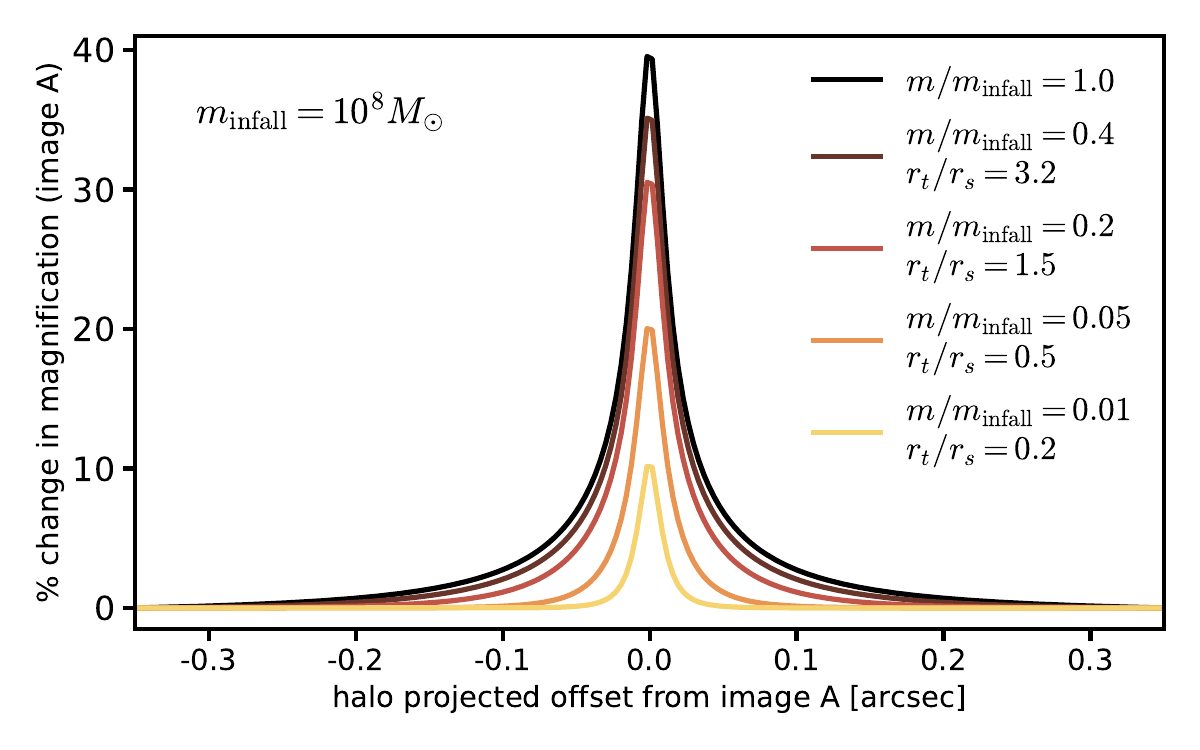}
    \includegraphics[trim=0cm 1cm 0cm
	0cm,width=0.48\textwidth]{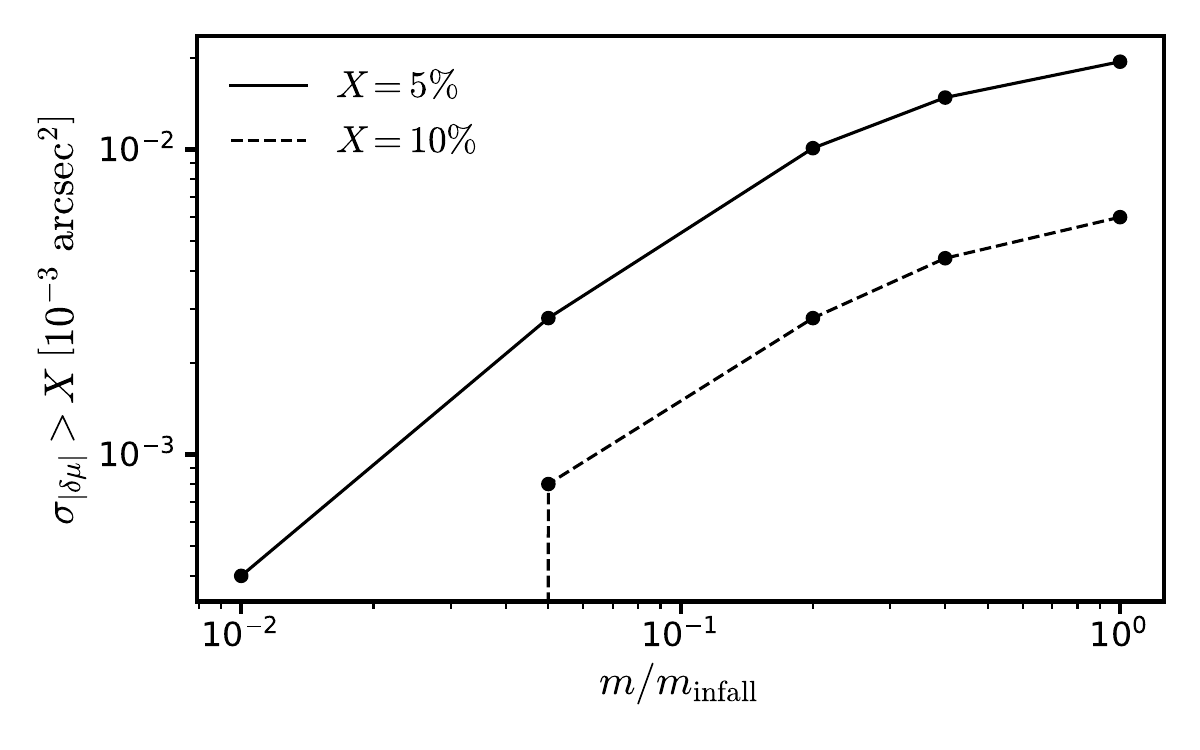}
	\caption{\label{fig:magcross}  {\bf{Left:}} the 1D magnification cross section of a halo as a function of its mass after tidal stripping. Curves show the perturbation to an image magnification as a function of the halo's projected offset from image A (see Fig.~\ref{fig:mockimage}), and colors represent varying degrees of tidal stripping, as indicated by the ratio $m/m_{\rm{infall}}$ and the truncation radius in units of $r_{\rm s}$. Here $m$ represents the mass of the subhalo that remains inside a sphere of radius $r_{\rm{200}}$, where $r_{\rm{200}}$ is evaluated at infall. {\bf{Right}}: the cross section $\sigma_{|\delta \mu|}$ in units $\rm{arcsec^2}$ for producing a change in image magnification greater than $5 \%$ (solid lines) and $10 \%$ (dashed lines) as a function of the subhalo mass after tidal stripping.}
\end{figure*}

In Fig.~\ref{fig:magcross} we show the magnification cross section for a subhalo with a mass at infall of $10^8 M_{\odot}$ that has experienced various degrees of tidal stripping. The colored curves represent the perturbation caused by a subhalo if it is stripped to $40\%$, $20\%$, $5\%$, and $1\%$ of its infall mass. Based on the value of $m_{\rm{bound}} / m_{\rm{infall}}$, we use Eq.~\eqref{eqn:rhonfw} to calculate the density profile, and then ray trace through the lens system to the source plane to calculate the image magnification. We assume a background source size of $5$ parsecs. 

As shown by Fig.~\ref{fig:magcross}, even subhalos that have lost $80 \%$ of their mass since infall can produce a significant perturbation to the flux ratios. This reflects the fact that flux ratios are sensitive to the central density of a halo, and initially tidal stripping mainly removes material from $r > r_{\rm s}$. The right-hand panel, which shows the cross section for producing a change in image magnification greater than $5 \%$ (solid lines) and $10 \%$ (dashed lines) shows that the perturbation to the image magnification is not simply a function of the bound mass, but depends on how the mass is distributed inside the subhalo. From both of these figures, we can conclude that once the halo becomes strongly disrupted $m_{\rm{bound}} / m_{\rm{infall}} < 0.2$, the cross section for producing a detectable change in image magnification rapidly decreases.

From Fig.~\ref{fig:boundmassfunc}, we see that the bound mass function is suppressed by a factor of $\sim 20$ relative to the infall mass function---many subhalos have therefore lost $80 \%$ of their mass since infall. From Fig.~\ref{fig:magcross}, we expect their lensing signal to be strongly suppressed, and populations of subhalos subject to this degree of tidal stripping will lead to different flux ratio statistics. To investigate this quantitatively, we create a mock lens system and calculate the probability of measuring a given set of flux ratios given the relative image positions, $\boldsymbol{x}$, and the imaging data, $\boldsymbol{I}$, $p\left(f_{\rm{B/A}}, f_{\rm{C/A}}, f_{\rm{D/A}} | \boldsymbol{x},\boldsymbol{I}\right)$. Here, the imaging data refers to the spatially resolved emission from the lensed quasar host galaxy that appears as a magnified, extended lensed arc around the main deflector. We do not include substructure in the lens model when creating this mock lens, so the statistical imprint of substructure will add scatter around the median flux ratios of the mock lens. 

\begin{figure}
	\includegraphics[trim=2cm 0cm 1cm
	0cm,width=0.33\textwidth]{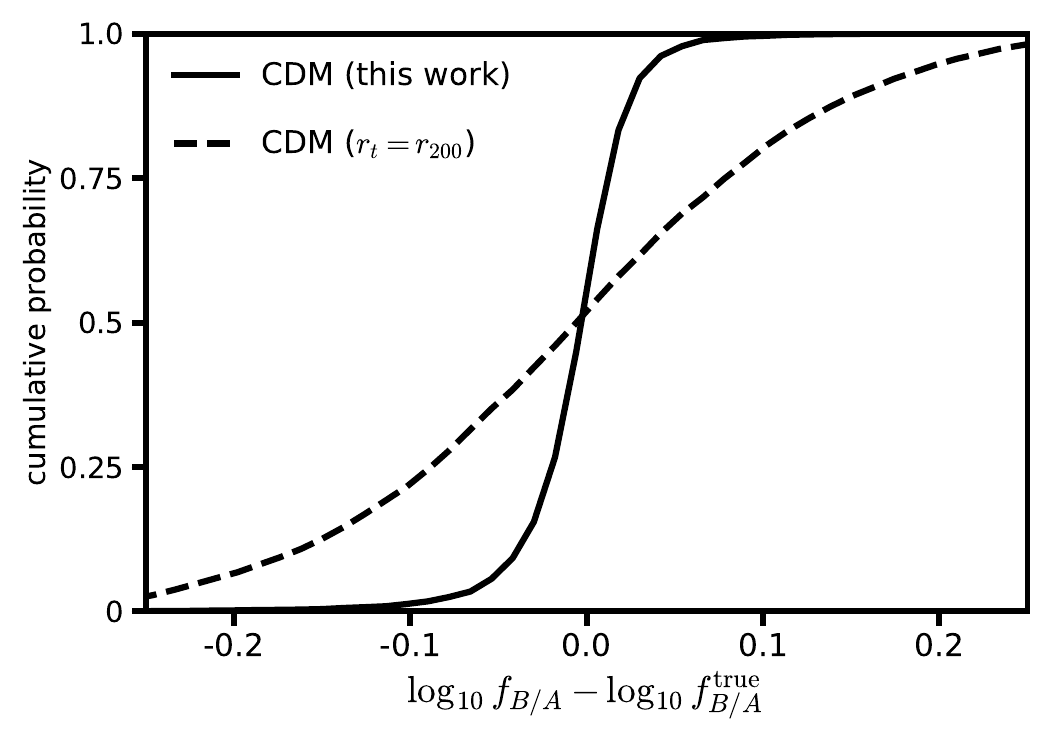}
    \includegraphics[trim=2cm 0cm 1cm
	0cm,width=0.33\textwidth]{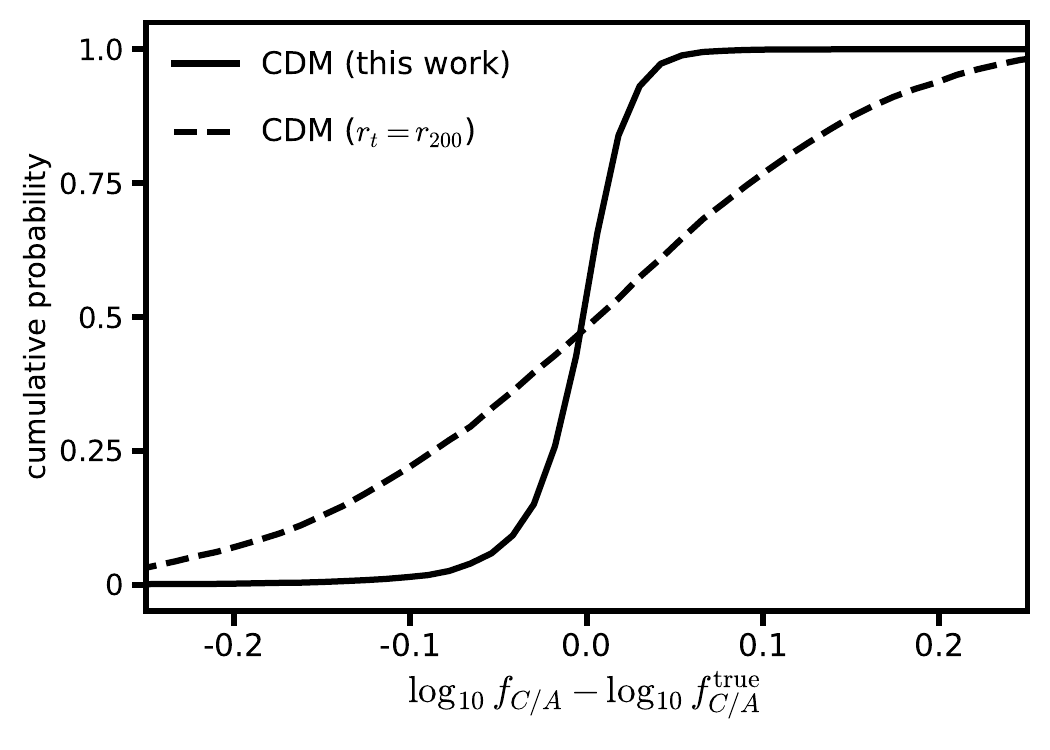}
    \includegraphics[trim=2cm 0cm 1cm
	0cm,width=0.33\textwidth]{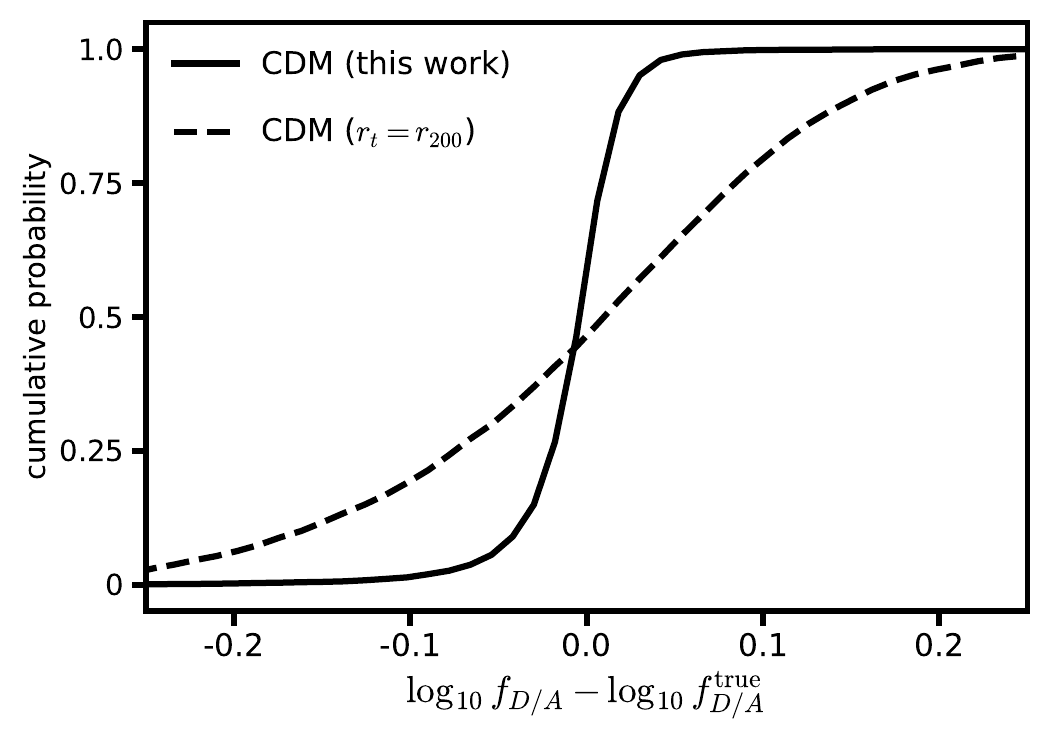}
	\caption{\label{fig:frcumulative} The cumulative distribution of flux ratio perturbations caused in CDM with the tidal stripping model presented in this work (solid black) and a model with no tidal stripping dashed black. The x axis shows the difference in the log flux ratios among the four images of the mock lens shown in Fig.~\ref{fig:mockimage}.}
\end{figure}

To compute $p\left(f_{\rm{B/A}}, f_{\rm{C/A}}, f_{\rm{D/A}} | \boldsymbol{x},\boldsymbol{I}\right)$, we simultaneously reconstruct the relative image positions, the surface brightness of the lensed arc, and the surface brightness of the lensed quasar host galaxy in the presence of a full population of dark matter subhalos. To reconstruct the imaging data and relative image positions with substructure we follow the lens modeling approach outlined by \cite{Gilman24}, using the same background spiral galaxy and source light model. To isolate the effects of substructure from uncertainties associated with the main deflectors mass profile, we have also used a simplified model of an isothermal elliptical power-law profile plus external shear in both the creation and the modeling of the mock lens. We omit line-of-sight halos and additional angular structure in the main deflector to isolate the effect of subhalos in the main lens plane from other sources of perturbation. 

We calculate the flux ratios for $10,000$ different subhalo populations created using {\tt{pyHalo}} with the tidal stripping model discussed in the previous section. The cumulative distributions of the flux ratios, relative to the ``true" flux ratio of the mock lens, are shown in Fig.~\ref{fig:frcumulative}. We have expressed the results as the $\log$ of the flux ratios so that each observation has the same scale. The black curves illustrate the distribution of flux ratios expected in cold dark matter, where the solid curve uses the tidal stripping model presented in this work, and the dashed curve adds a truncation at $r_{\rm{200}}$, effectively ignoring tidal stripping effects. From these distributions, we see that the model for tidal stripping presented in this work has important implications for interpreting flux ratio anomalies in lensed quasars. For example, a deviation of $20\%$ from the flux ratios predicted by a smooth lens model would be quite rare in CDM, given the tidal stripping model we have developed, but this level of perturbation occurs more frequently when tidal stripping is not accounted for. Constraints on dark matter properties that incorporate the improved model of tidal stripping presented in this work will be presented in a forthcoming publication. 

\section{Discussion}\label{sec:disscussion}
We have presented a model for tidal stripping of dark matter subhalos that predicts their bound masses and density profiles given their concentrations at infall, their infall redshifts, and the host halo concentration. The model is calibrated to accurately predict the bound mass fraction of subhalos that appear in projection within $30$ kpc of the host halo center, which coincides with the areas of dark matter halos where strong gravitational lenses can characterize the properties of dark substructure. We have implemented this model in the open-source software {\tt{pyHalo}}\footnote{https://github.com/dangilman/pyHalo} for future use in strong gravitational lensing analyses. Our main conclusions can be summarized as follows: 
\begin{enumerate}[label=(\roman*)]
\item The majority ($\sim 87\%$) of subhalos that appear near the Einstein radius in projection have lost in excess of $\sim 90 \%$ of their mass since infall due to tidal stripping, on average. Only $\sim 7 \%$ of subhalos that appear near the Einstein radius retain more than $20 \%$ of their mass since infall. 
\item The amplitude of the bound mass function is suppressed relative to the infall mass function by a factor of $20$, on average. This results in a projected mass in dark subhalos of $3.2 \pm 0.6 \times 10^6 M_{\odot}\,\rm{kpc^{-2}}$ in the mass range $10^8$--$10^{10} M_{\odot}$. 
\item The median mass loss experienced by subhalos is a strong function of their infall times and concentrations, with more concentrated subhalos at infall losing less mass on average than less concentrated subhalos. Subhalos that appear near the Einstein radius which have retained most of their mass since infall are therefore either significantly more concentrated than average, were accreted onto the host more recently than $\sim 2 \ \rm{Gyr}$, have wide orbits with large pericenters such that they do not experience strong tidal forces, or a combination of these factors. 
\end{enumerate}

Previous strong lensing analyses with flux ratios have used an implementation for tidal stripping that is significantly simplified compared to the model we have presented in this work. For instance, the authors of \cite{Gilman20} used a model for halo truncation that predicted $r_{\rm t}/r_{\rm s} \sim 1$--$5$ for most halos, which corresponds to mass loss $\sim 50\%$--$80 \%$, on average. More recently, \cite{Keeley24} presented constraints on warm dark matter using an earlier version of the tidal stripping model we have developed in this work. The model used by \cite{Keeley24} implemented less aggressive tidal stripping, predicting a bound mass function suppressed relative to the infall mass function by a factor $7$, with the difference primarily stemming from the treatment of indirect and direct infall distributions for subhalos and sub-subhalos. The broad priors on the normalization of the subhalo mass function used by both of these works were chosen in part to account for uncertainties associated with the tidal evolution model. Moving forward, using the model presented in this paper, we have a more reliable prediction for the number of subhalos expected to perturb image flux ratios in CDM, how the populations of subhalos present in strong lenses depend on their time and concentration at infall, and how the tidal evolution affects their density profile. We will use the tidal stripping framework discussed in this work in forthcoming publications that analyze the properties of $\sim 30$ quadruply imaged quasars recently observed through a Cycle 1 James Webb Space Telescope (JWST) program GO-2046 \cite{Nierenberg24}.

A criterion that must be satisfied to use this model in other cosmological scenarios is that the evolution of the host halo remains unchanged relative to the model for the host halo evolution used in this work. As the host halo is a group-scale object $m_{\rm{200}} \sim 10^{13} M_{\odot}$, this requirement effectively means we can use this tidal stripping model for dark matter theories and primordial power spectra that deviate from $\Lambda$CDM only on subgalactic scales. For example, warm dark matter models with small-scale suppression of the matter power spectrum leave the properties of very massive group-scale hosts unchanged (see Appendix \ref{sec:SHMF_WDM}), but result in a strong suppression of the concentration-mass relation for low-mass objects, e.g.,~\cite{Ludlow16}. Our model predicts (see Fig.~\ref{fig:mboundwdm}) that halos with lower infall concentrations will be more susceptible to tidal stripping than their CDM counterparts. Conversely, an enhancement of the primordial matter power spectrum \cite{Gilman22,Dekker:2024nkb,Esteban24,We24}, or primordial non-Gaussianity \cite{Stahl23}, on scales $k > 10\,\rm{Mpc^{-1}}$ leads to significantly higher concentrations for low-mass halos without significantly affecting the internal structure of group-scale hosts. With higher infall concentrations, our model predicts subhalos become more resilient to tidal stripping, increasing their lensing efficiency and possibly allowing them to retain more luminous matter inside a tidal field. The model we have presented may not provide as robust predictions for dark matter models in which additional processes besides gravity impact the evolution of halo structure. For example, in self-interacting dark matter halos go through a period of core formation followed by core collapse (see Ref \cite{Adhikari22}, and references therein). These processes may affect the tidal evolution of subhalos in a way our model does not fully capture.  

Historically, the tidal evolution of subhalos around a host has been regarded as a primary source of systematic uncertainty for strong lensing analyses. Quantifying what transpires after a halo becomes a subhalo requires expensive numerical simulations, and accounting for these effects in Bayesian inferences requires computationally expensive forward modeling. The techniques we have presented in this paper transform this limiting source of systematic uncertainty into an additional signal with which to constrain the nature of dark matter and the initial conditions for structure formation. Moving into a new era of large surveys expected to discover thousands of strong lens systems, models such as the one presented here are necessary to maximize the scientific return of the experiments and lead to more reliable inferences of dark matter physics from cosmological observations. 

\section*{Acknowledgements}

X.D. and T.T. acknowledge support from the National Science Foundation through Grants No. NSF-AST-1836016 and No. NSF-AST-2205100, by the Gordon and Betty Moore Foundation through Grant No. 8548, and by the National Aeronautics and Space Administration through Grants No. JWST-GO-2046. D.G. acknowledges support from the Brinson Foundation provided through a Brinson Prize Fellowship grant. This research was supported in part by Grants No. NSF PHY-2309135 to the Kavli Institute for Theoretical Physics (KITP). 

This work used computational and storage services associated with the Hoffman2 Cluster which is operated by the UCLA Office of Advanced Research Computing’s Research Technology Group.

\section*{Data Availability}
The data that support the findings of this article are openly available~\cite{du_2025_15005098}.

\appendix

\section{Subhalo evolution in WDM host potential}
\label{sec:SHMF_WDM}

\begin{figure}
\includegraphics[width=0.45\textwidth]{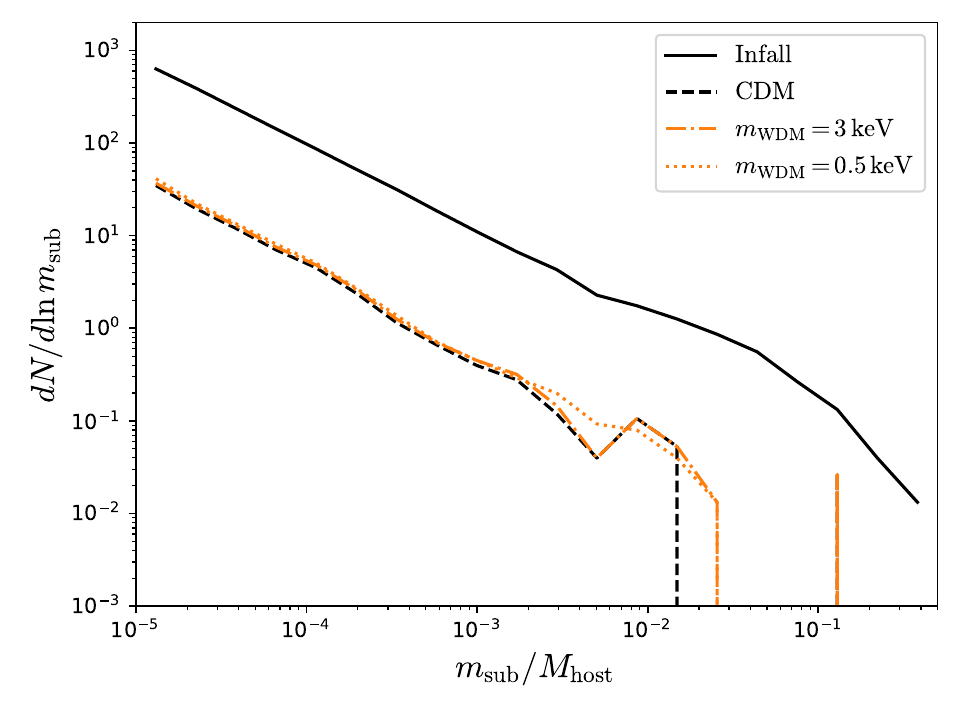}
\caption{Mass function of subhalo evolved in WDM hosts compared with that in a CDM host. Here we have assumed the same infall mass function for all cases. Only subhalos within a projected distance of $30\,{\mathrm{kpc}}$ are included.}
\label{fig:SHMF_CDM_WDM}
\end{figure}

Figure~\ref{fig:SHMF_CDM_WDM} shows the bound mass function of subhalos evolved in WDM hosts (dot-dashed and dotted curves) compared with that in a CDM host (black dashed curve) from idealized {\sc galacticus} simulations, assuming the same infall mass function (solid green curve). Little difference is seen in the bound mass function even for the case $m_{\mathrm{WDM}}=0.5\,{\mathrm{keV}}$.

\section{Mass functions of directly and indirectly infalling subhalos}
\label{app:zinfall}

\begin{figure}
\includegraphics[width=0.45\textwidth]{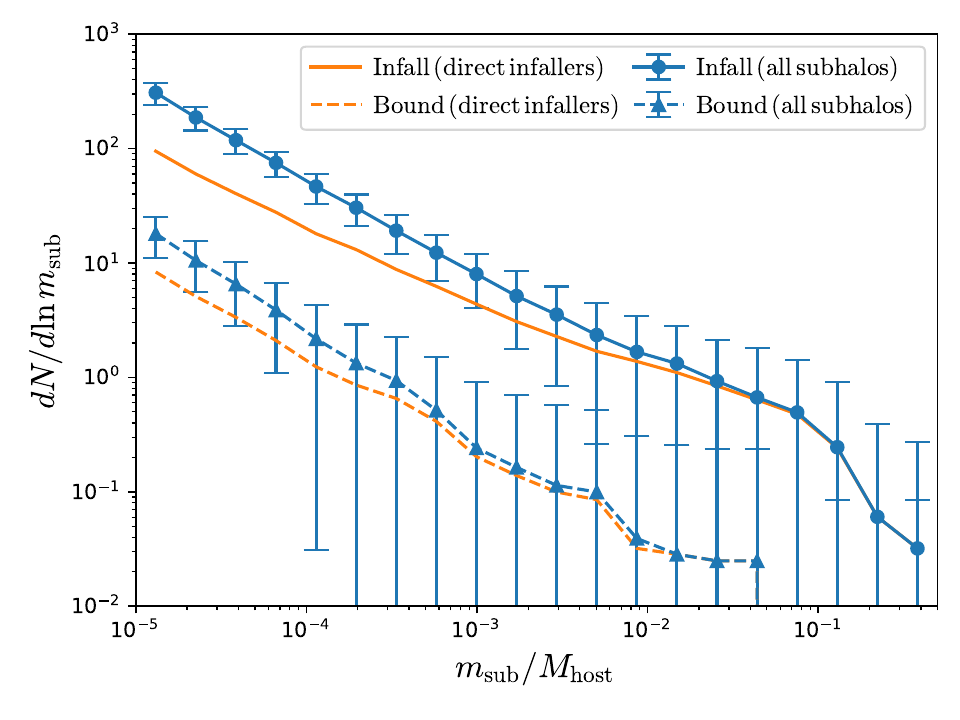}
\caption{Mass function of direct infallers (orange curves) compared with that of all subhalos (blue curves with error bars). The solid curves are obtained using the infall mass while the dashed curves are obtained using the bound mass. Only subhalos within a projected distance of $30\,{\mathrm{kpc}}$ are included.}
\label{fig:SHMF_direct_indirec_com}
\end{figure}

At high mass ratios, the subhalos are dominated by those that directly fell into the host (see Fig.~\ref{fig:SHMF_direct_indirec_com}). However, as the mass ratio decreases, there are more and more indirectly infalling subhalos. At $m_{\mathrm{sub,bound}}/M_{\mathrm{host}}\sim 10^{-5}$, only $46\%$ of subhalos are direct infallers. So it is important to choose appropriate effective infall redshifts for those indirect infallers if we treat all subhalos as direct infallers.

\section{Tidal tracks}
\label{app:track}

\begin{figure*}
\includegraphics[width=0.45\textwidth]{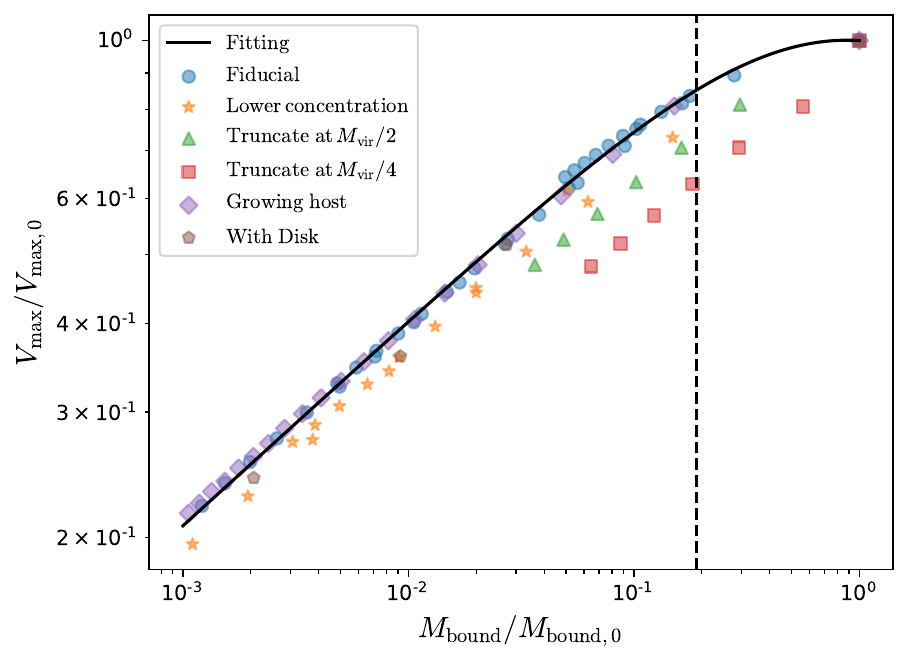}
\includegraphics[width=0.45\textwidth]{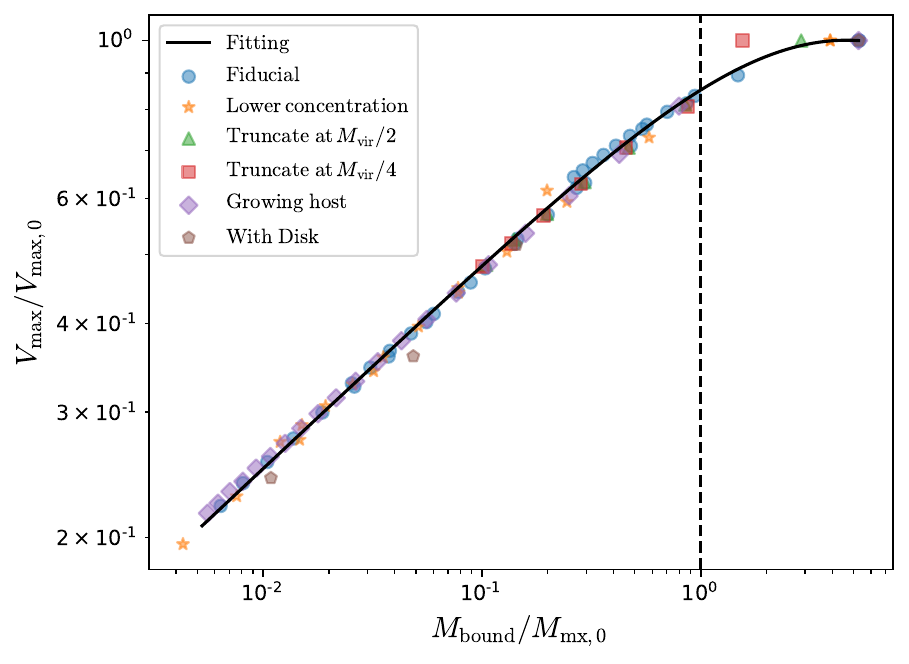}
\caption{Maximum circular velocity with respect to the bound mass of subhalos. Left panel: the bound mass is shown in units of the initial bound mass, i.e., the infall mass. Right panel: the bound mass is shown in units of $M_{\rm mx}=M(r<R_{\rm max})$ with $R_{\rm max}$ the radius at which the maximum circular velocity is reached. The dots, stars, diamonds, and pentagons are from Ref.~\cite{Du:2024sbt}.}
\label{fig:vmax}
\end{figure*}

\begin{figure*}
\includegraphics[width=0.45\textwidth]{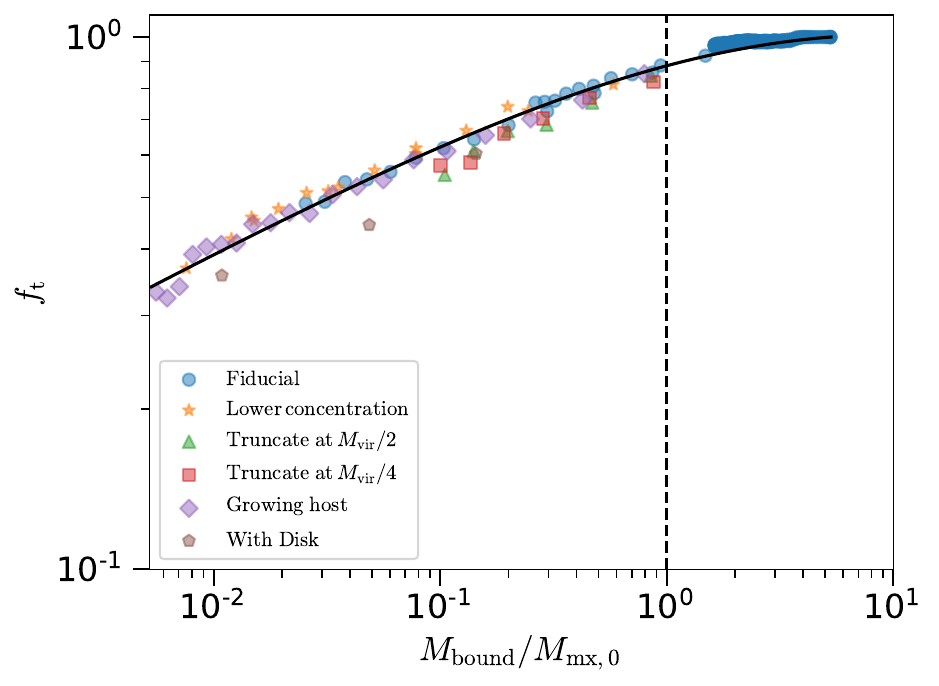}
\includegraphics[width=0.45\textwidth]{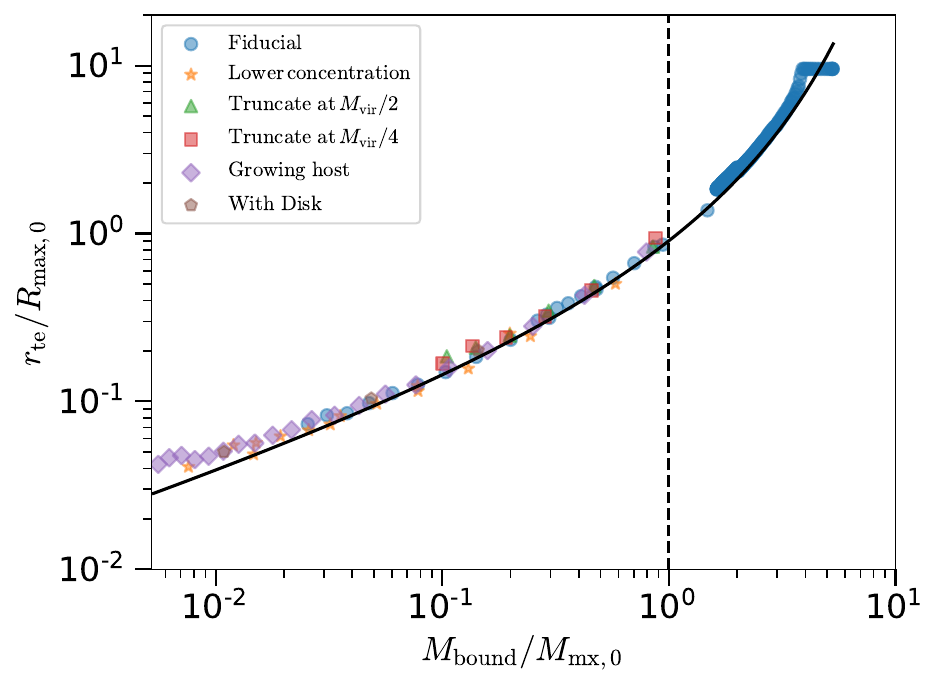}
\caption{The normalization parameter $f_{\rm t}$ and truncation radius $r_{\rm t}$ in the density transfer function Eq.~\eqref{eq:rho_T} as a function of bound mass. The dots, stars, diamonds, and pentagons are from Ref.~\cite{Du:2024sbt}.}
\label{fig:ft_rt}
\end{figure*}

In this work, we make use of the fitting tidal tracks from Ref.~\cite{Du:2024sbt}, in which fitting functions for the maximum circular velocity $V_{\mathrm{max}}$, and density transfer function are derived in terms of the bound mass fraction $M_{\mathrm{bound}}/M_{\mathrm{ bound,0}}$. However, these functions are fitted to idealized noncosmological simulations. The initial bound mass at infall $M_{\mathrm{bound,0}}$ is defined based on the matter density of the Universe at $z=0$, thus preventing us from directly using the tidal tracks for our purpose. For example, considering a halo with an NFW profile, the virial mass will be different, according to the definition $\overline{\rho}(<R_{\mathrm{vir}})=\Delta_{\rm vir}\rho_{\mathrm{vir}}(z)$, if the halo is at a different redshift larger than $0$. Equivalently, this is to truncate the density profile at a smaller radius. We have run two idealized N-body simulations as in Ref.~\cite{Du:2024sbt} with NFW initial halo profiles, but truncated the density profiles at $M_{\mathrm{vir}}/2$ and $M_{\mathrm{vir}}/4$. As shown in the left panel of Fig.~\ref{fig:vmax}, the ratio $V_{\mathrm{max}}/V_{\mathrm{max},0}$ is lower than the fitting curve derived in Ref.~\cite{Du:2024sbt} (black curve). But if we plot $V_{\mathrm{max}}/V_{\mathrm{max},0}$ with respect to $M_{\mathrm{bound}}/M_{\mathrm{mx,0}}$, we do get a universal track for all cases. Here $M_{\mathrm{mx},0}$ is the initial enclosed mass with $R_{\rm max,0}$ (see also \cite{Errani:2021rzi}). Thus in this work, we will do some conversions before applying the fitting tidal tracks from Ref.~\cite{Du:2024sbt}. Figure~\ref{fig:ft_rt} shows the normalization parameter $f_{\rm t}$ and truncation radius $r_{\rm t}$ in the density transfer function Eq.~\eqref{eq:rho_T} as a function of $M_{\mathrm{bound}}/M_{\mathrm{mx,0}}$. Similarly, we get universal tracks for all different simulations.

\clearpage
\bibliography{bibliography}

\end{document}